\documentclass[pra,11pt,tightenlines,onecolumn,superscriptaddress]{revtex4}
\usepackage{color}
\usepackage{lineno,hyperref}
\usepackage{mathrsfs}
\usepackage{amssymb}
\usepackage{amsthm}
\usepackage{todonotes}
\usepackage{enumitem}
\usepackage{amsmath}
\usepackage{units}
\usepackage[capitalise]{cleveref}
\usepackage{dsfont}
\usepackage{bm}
\usepackage{bbm}
\usepackage{xcolor}
\usepackage{subfig}
\usepackage[]{graphicx}

\usepackage{physics}
\usepackage{comment}

\usepackage{natbib} 

\newtheorem{remark}{Remark}

\newtheorem{fact}{Fact}

\DeclareMathOperator{\sgn}{sgn}
\DeclareMathOperator{\arctanh}{arctanh}
\newcommand{\new}[1]{\textcolor{black}{{#1}}}

\begin{document}
\title{%
Channel Discord and Distortion}
\author{Wei-Wei Zhang}\email{zhangweiwei2021@gusulab.ac.cn}
\address{Max Planck Institute for the Science of Light, Staudtstraße 2, 91058 Erlangen, Germany}
\address{Institute for Theoretical Physics II,  Friedrich-Alexander-Universit{\"a}t Erlangen-N{\"u}rnberg,  Staudtstraße 7, 91058 Erlangen, Germany}
\address{Centre for Engineered Quantum Systems,
    School of Physics, The University of Sydney,
    Sydney, Australia}
\author{Yuval R.\ Sanders}
\address{Department of Physics and Astronomy and ARC Centre of Excellence for Engineered Quantum Systems, Macquarie University, Sydney, Australia}
\author{Barry C.\ Sanders}
\address{Shanghai Branch, National Center for Physical Sciences at Microscale,
    University of Science and Technology of China,
    Shanghai 201315, China}
\address{Institute for Quantum Science and Technology, University of Calgary, Alberta, Canada T2N 1N4}

\begin{abstract}
Discord, 
originally notable as a signature of bipartite quantum correlation,
in fact can be nonzero classically,
i.e., arising from noisy measurements by one of the two parties.
Here we redefine classical discord to quantify channel distortion,
in contrast to the previous restriction of classical discord to a state,
and we then show a monotonic relationship between classical (channel) discord and channel distortion.
We show that classical discord is equivalent to 
(doubly stochastic)
channel distortion
by numerically discovering a monotonic relation between discord and total-variation distance
for a bipartite protocol with one party having a noiseless channel and the other party having a noisy channel. 
Our numerical method includes randomly generating doubly stochastic matrices for noisy channels and averaging over a uniform measure of input messages. 
Connecting discord with distortion establishes discord as a signature of classical, not quantum,  channel distortion.
\end{abstract}
\maketitle
\tableofcontents
\section{Introduction}
Discord is often touted as a quantifier of quantum correlations in a state,
with nonzero discord said to imply that observed correlations
transcends
non-quantum
(i.e., `classical')
limits~\cite{HV01,OZ01},
akin to,
but different from, 
a Bell inequality~\cite{brunner2014bell}.  Treated as a quantum resource operationalized by state merging~\cite{Horodecki2007},
discord in quantum computing protocols~\cite{Lanyon2008, datta2008} is believed by many to deliver a quantum advantage to some protocols~\cite{Merali2011}.
However, this quantum nature of discord has been challenged by stochastic information,
which shows that discord is due to noisy measurement by one of the two parties~\cite{GdOS15}.
Essentially, discord can be understood in terms of a protocol amenable
to a stochastic-information interpretation,
and this interpretation
fails if and only if (iff)
the two parties share bipartite entanglement~\cite{GdOS15}.
Discord thus serves as a fascinating starting point for studying stochastic information.

Previous work analyzed state discord in the context of classical states~\cite{GdOS15},
i.e., analyzing discord as a signature of classical rather than quantum correlations.
Here we introduce the concept of classical channel discord,
which is based on averaging over all allowed channel input states for a given channel,
with the previous definition used in assessing how much discord is added to a state by the given channel.
Our expression for channel discord
is amenable to numerical evaluation,
which shows channel discord is monotonic with respect to channel distortion.
We augment this numerical analysis by solving analytically the small but nontrivial two-bit case (each of two parties holds one bit) to confirm our numerics for this case and establish a path for proving discord-distortion monotonicity,
which is a challenging calculation as we show.
Our monotonicity result establishes meaningfully that channel discord and distortion are essentially equivalent.

Mathematically, discord in a two parties state refers to an apparent discrepancy between two expressions for mutual information obtained by two parties
named here as Alice~(A) and Bob~(B),
one quantity depending on joint probability and then other depending on conditional probability.
The usual view of the reason for quantum discord in a state is that conditional information must be adapted to the quantum case by introducing measurement,
and this incompatibility gives rise to nonzero discord~\cite{OZ01}.
Quantum discord in a state has been shown to be equivalent to \new{classical}  discord if and only if entanglement between the two parties is zero,
and this equivalence has been explained by showing that,
in the absence of entanglement,
discord represents one party,
namely Bob,
suffering from noisy measurement whereas Alice's measurements are ideal~\cite{GdOS15}. 
Our goal is to show that classical channel discord,
obtained by averaging channel-added discord to all allowed channel input states,
is equivalent to channel distortion
by establishing
a monotonic relation between discord and total-variation,
or Kolmogorov, distance,
which quantifies channel distortion.
We analyze the general case numerically and solve analytically for the two-bit case.

Our article is structured as follows. 
In~\S\ref{sec:background} we summarise essential background 
on stochastic information, discord,  
total-variation distance
and doubly stochastic channels.
Our approach is described in~\S\ref{sec:approach}
and our approach specifically
elaborates on our model for describing a noisy protocol for creating channel discord.
Furthermore, \S\ref{sec:approach}
presents our notation and mathematical expressions,
and our methods for solving these expressions numerically.
Subsequently,
in~\S\ref{sec:results},
we present our numerical results and explain the plots,
and we discuss the results thoroughly in~\S\ref{sec:discussion}.
In~\S\ref{sec:conclusions},
we summarise our claims and provide an outlook.
In Appendices~\ref{sec:probvec} and~\ref{sec:hadamardcalculus} we introduce convenient notation for probability vectors and what we call Hadamard calculus  respectively,
which fundamental to our approach.

\section{Background }
\label{sec:background}
In this section we discuss the background and context for our work.
In~\S\ref{subsec:stochasticinformation}
we discuss informational states including what we call stochastic information,
which is a probabilistic mixture of definite informational states;
specifics regarding probabilistic information states are explained in Appendix~\ref{sec:probvec},
which is based on the Hadamard notation explained in Appendix~\ref{sec:hadamardcalculus}.
Included in this discussion,
we review the notions of entropy by considering shared information between parties, 
and we also review the notions of entropy and mutual information,
all in the elegant Hadamard notation
elaborated in Appendix~\ref{sec:hadamardcalculus},
which we apply for the first time to this application.
Then,
in~\S\ref{sec:stochasticmap},
we explain mappings of information states in terms of channels with special emphasis on doubly  stochastic channels.
In this subsection we discuss ways to quantify how stochastic a channel is.
Finally,
in~\S\ref{subsec:statediscord},
we review the notion of classical  discord for states
and the concepts of total-variation distance
for stochastic information states;
quantum discord is explained in Appendix~\ref{sec:quantumdiscord}.

\subsection{Stochastic Information}
\label{subsec:stochasticinformation}
In this subsection
we review the concept and mathematical framework for stochastic information based on the probability-vector representation elaborated in Appendix~\ref{sec:probvec} which uses  Hadamard calculus introduced in Appendix~\ref{sec:hadamardcalculus}.
Then
we discuss known concepts concerning entropy of a stochastic-information state by using Hadamard calculus introduced  in Appendix~\ref{sec:hadamardcalculus}.
Finally,
we review bipartite stochastic information including conditional entropy and mutual information.


The joint probability of messages shared between Alice,
whose message size is~$M^\text{A}$,
and Bob,
whose message size is~$M^\text{B}$,
is the bipartite matrix
\begin{equation}
\label{eq:pAB}
\bm{p}^{\text{AB}}
    =\sum_{mm'}
        p^{\text{AB}}_{mm'}
        \bm{\delta}^\text{AB}_{mm'}
        \in\text{mat}_{M^\text{A}\times {M}^\text{B}}\left(\mathbb{R}^{\geq0}\right), 
   \norm{\bm{p}^{\text{AB}}}=1,
\end{equation}
using the notation that $\text{mat}_{M^\text{A}\times M^\text{B}}\left(R\right)$
refers to matrices with~$M^\text{A}$ rows and~$M^\text{B}$ columns whose entries are from any ring~$R$.
The norm is defined by Eq.~(\ref{eq:norm}).
Here we have let~$\bm\delta^{\text{AB}}_{mm'}$ denote a versor for message~$m'$ as discussed in Appendix~\ref{sec:probvec}.

In concordance with quantum-information nomenclature,
we refer to~$\bm{\delta}_{\check{m}}$ as a `pure state'~\cite{nielsen2002quantum}. 
Impure states refer to `mixed' states,
which are probabilistic mixtures of pure states
and are obtained as a probabilistic mixture of any pure state.
We discuss mixed and pure bipartite stochastic-information states in Appendix~\ref{sec:probvec}.

Now we discuss the entropy of the probability vector representing the mixed message.
Mixedness of a state~$\bm{p}\in\mathbb{R}^M$
is quantified by entropy~\cite{shannon1948mathematical}
\begin{equation}
\label{eq:H}
    0\leq H(\bm{p})
    :=-\bm{p}\odot\log\bm{p}\leq \log M
\end{equation}
using Hadamard notation explicated in Appendix~\ref{sec:hadamardcalculus}.
A state is pure iff its entropy is zero,
which follows from
$\bm{p}\circ\log\bm{p}=\bm0$
for a versor.
The lower bound for the entropy~(\ref{eq:H})
is $H(\bm{p})=0$ for a pure state.
The upper bound for entropy is $H(\bm{p})=\log M$ 
for a uniformly mixed state,
with $M$ as the size of $\bm{p}$.
A high-entropy state is a state whose entropy is close to this bound.
The joint state~(\ref{eq:pAB}) has
joint entropy $H^\text{AB}$~(\ref{eq:H})
and total message size~$M$.
In our analysis, we always assume,
without loss of generality,
that
\begin{equation}
\label{eq:MAMBM}
    M^\text{A}\equiv M^\text{B}
        \implies M^\text{A}=\sqrt{M}=M^\text{B}.
\end{equation}
Thus, the joint entropy of the bipartite versor~(\ref{eq:bipartiteversor}) is zero as required for a pure state.
The matrix representation of~$\bm{p}^{\text{AB}}$
is an $M^\text{A}\times M^\text{B}$ matrix,
with nonnegative real entries such that the sum of all entries is one.

The marginal distribution is obtained by ignoring the other party's share of the mixed state.
Hence,
Alice's marginal distribution is
the probability vector
\begin{equation}
\label{eq:marginalprob}
    \bm{p}^\text{A}
        :=\left(\sum_{m'}p^\text{AB}_{mm'}\right),\;
    \norm{\bm{p}^\text{A}}
        =\sum_m p^\text{A}_m
\end{equation}
using the unit one-norm.
Similarly,
we construct the marginal distribution $\bm{p}^\text{B}$
by summing over Alice's degree of freedom.

Alice's state conditioned on Bob's state is
\begin{equation}
\label{eq:pA|B}
\bm{p}^{\text{A}|\text{B}}
    :=\bm{p}^\text{AB}\oslash\bm{p}^\text{B}
    =\left(\frac{p^\text{AB}_{mm'}}
        {p^\text{B}_{m'}}\right)
\end{equation}
with~$\oslash$ explained in Appendix~\ref{sec:hadamardcalculus}.
The last term of Eq.~(\ref{eq:pA|B})
displays row-vector elements
$\bm{p}^\text{A}_m$
obtained by element-wise division of each matrix element $p^\text{AB}_{mm'}$
by respective column-vector elements~$p^\text{B}_{m'}$.
Similarly, the conditional probability distribution
for Bob
is $\bm{p}^{\text{B}|\text{A}}
    =\bm{p}^\text{AB}\oslash\bm{p}_\text{A}$
analogous to~(\ref{eq:pA|B}).

The entropy of the conditional probability distribution $\bm{p}^{\text{A}|\text{B}}$
is
\begin{equation}
    H^{\text{A}|\text{B}}\left(\bm{p}^\text{AB}\right)
      =H\left(\bm{p}^{\text{A}|\text{B}}\right).
\end{equation}
\begin{fact}
\label{fact:condpure}
A bipartite stochastic state
$\bm{p}^\text{AB}$,
which decomposes to $\bm{p}^\text{AB}\oslash\bm{p}^\text{B}$
and to $\bm{p}^\text{AB}\oslash\bm{p}^\text{A}$,
is conditionally pure
iff
\begin{equation}
\label{eq:condpure}
    H^{\text{A}|\text{B}}
        \equiv0
        \equiv H^{\text{B}|\text{A}}.
\end{equation}
\end{fact}
Consequently,
$\bm{p}^\text{AB}$
is conditionally pure iff it is permutationally equivalent to a diagonal matrix;
i.e., in diag$([0,1])$,
which refers to the set of diagonal matrices whose entries are each in the real-number interval~$[0,1]$.
Thus,
\begin{equation}
\label{eq:cpdiagonal}
    \exists\sigma\in S_{M^\text{A}},\sigma'\in S_{M^\text{B}}:
    \Pi_\sigma\bm{p}^\text{AB}\Pi_{\sigma'}
        \in\text{diag}_{\min\{M^\text{A},M^\text{B}\}}([0,1]),
\end{equation}
i.e., is diagonal of size $\min\{M^\text{A},M^\text{B}\}
\times\min\{M^\text{A},M^\text{B}\}$.
Furthermore,
conditional probability distributions  $\bm{p}^{\text{A}|\text{B}}$ and $\bm{p}^{\text{B}|\text{A}}$,
which are obtained from a bipartite stochastic pure information state $\bm{p}^\text{AB}$,
are necessarily pure.

Operationally speaking,
a bipartite state is conditionally pure only if Alice's pure state can be known by Bob 
after he measures his share of the joint stochastic-information state 
and vice versa.
Consequently,
the bipartite versor~(\ref{eq:bipartiteversor})
used in Eq.~(\ref{eq:pAB})
is conditionally pure.

Mutual information
\begin{align}
\label{eq:I}
    I^{\text{A};\text{B}}\left(\bm{p}^\text{AB}\right)
     :=&H^\text{A}\left(\bm{p}^\text{AB}\right)
        +H^\text{B}\left(\bm{p}^\text{AB}\right)
        -H^\text{AB}\left(\bm{p}^\text{AB}\right)\nonumber\\
    =&\bm{p}^\text{AB}
    \odot\log
    \left(\bm{p}^\text{AB}\oslash\bm{p}_\text{A}\oslash\bm{p}^\text{B}\right)
\end{align}
quantifies correlation between two parties, Alice and Bob,
with the last part of this expression expressed in a novel way by using Hadamard arithmetic.
An equivalent,
alternative mutual information definition is
\begin{equation}
\label{eq:J}
    J^\text{A;B}\left(\bm{p}^\text{AB}\right)
        :=H^\text{A}\left(\bm{p}^\text{AB}\right)
            -H^{\text{A}|\text{B}}\left(\bm{p}^\text{AB}\right)
        =I^{\text{A};\text{B}}\left(\bm{p}^\text{AB}\right).
\end{equation}
Consequently, 
\begin{equation}
\label{eq:classicaldiscorddef}
\Delta^{\text{A};\text{B}}
    \left(\bm{p}^\text{AB}\right)
    :=I^{\text{A;B}}\left(\bm{p}^\text{AB}\right)
        -J^\text{A;B}\left(\bm{p}^\text{AB}\right)
    \equiv0
\end{equation}
so $I^\text{A;B}\left(\bm{p}^\text{AB}\right)$~(\ref{eq:I})
and $J^\text{A;B}\left(\bm{p}^\text{AB}\right)$~(\ref{eq:J})
are equal.


\subsection{Stochastic map and stochastic matrix}
\label{sec:stochasticmap}
A noisy channel is any mapping that changes the entropy (or noise,
which is monotonically related) of a state in a non-decreasing way and adds noise to at least one state~\cite{cover2012elements}. 
We are specifically interested in noisy channels that can be represented as stochastic matrices that map probability vectors representing states~\cite{GdOS15}.

Under the action of a channel represented by matrix~$\mathcal E$,
the state,
represented by~$\bm{p}$,
maps to $\mathcal{E}\bm{p}$.
We require that~$\mathcal{E}$
is a square matrix with nonnegative entries such that either rows or columns sum to one.
Hence, the norm of the state~$\bm{p}$
is unchanged by the stochastic map by stochastic matrix~$\mathcal{E}$.
The entropy of this state after passing through the channel is $H(\mathcal{E}\bm{p})\geq H(\bm{p})$.
A doubly stochastic matrix is a stochastic matrix whose rows and columns both sum to one.

In the bipartite setting,
an identity mapping by Alice concomitant with a stochastic map~$\mathcal{E}$ by Bob,
yields the resultant bipartite state
\begin{equation}
\label{eq:pAB-1MpAB}
\bm{p}^\text{AB}
    \mapsto\mathbb{I}
        \bm{p}^\text{AB}\mathcal{E}
    =\bm{p}^\text{AB}\mathcal{E}.
\end{equation}
with the trivial identity map~$\mathbb{I}$ on Alice's side
and the noise matrix~$\mathcal{E}$
only acting on Bob's share.
By the Perron-Frobenius theorem,
$\mathcal{E}$ being stochastic or doubly stochastic
implies that  this mapping has at least one stationary vector with all entries being positive real numbers
with this vector corresponding to the largest eigenvalue of the matrix representing the mapping~\cite{milonni1976}.

Although we discuss discord and total-variation distance in terms of measurement described by a noisy measurement channel represented by a stochastic matrix,
we focus on doubly stochastic matrices due to the abundance of mathematical properties that we can exploit for generating and understanding our results.
For stochastic information theory,
doubly stochastic matrices are the non-quantum analogue of quantum completely positive trace-preserving maps~\cite{Gasbarri2018}.

Birkhoff's Theorem says that any doubly stochastic matrix can be written as a convex hull of permutation matrices,
which is known as the Birkhoff polytope~\cite{budish2009implementing}.
The doubly stochastic matrix thus represents a random permutation of bits in the string.
Furthermore,
for each strictly positive matrix $A$,
exactly one doubly stochastic matrix $T_A$ exists such that $T_A=DAD'$ with the diagonal matrices~$D$ and~$D'$  having positive diagonal-elements and themselves unique up to a scalar factor~\cite{sinkhorn1964relationship,johnson1981row}.
Here we present key background information on doubly stochastic matrices needed for our study. 
Specifically,
we define doubly stochastic matrices and connect these matrices with the Birkhoff polytope,
also known as a permutahedron.
A permutation~$\sigma$
is represented by a permutation matrix~$\Pi_\sigma$, 
whose entries are all zeroes and ones such that only one instance of one appears in each row or column.
For~$\bm\wp$ a length-$M!$ probability vector,
a permutahedron~$\mathcal E$ is the convex sum
\begin{equation}
\label{eq:permutahedron}
\mathcal{E}=\sum_{\sigma\in S_{M!}}\bm\wp\cdot\bm\Pi_\sigma,\;
\bm\Pi_\sigma\in\text{mat}_{M!}\left(\{0,1\}\right)
\implies\mathcal{E}\in\text{mat}_{M!}\left(\mathbb{R}\right).
\end{equation}
Note that $\mathcal{E}\Pi_\sigma=\Pi_\sigma\mathcal{E}$.

\subsection{Classical discord and distortion for a state}
\label{subsec:statediscord}
In Appendix~\ref{sec:quantumdiscord},
we summarise quantum discord;
in this subsubsection,
we summarise{ classical discord for stochastic-information states~\cite{GdOS15}.
Whereas quantum discord is the discrepancy~$\Delta^{\text{A};\text{B}}$ between mutual information $I$~(\ref{eq:I}) and $J$~(\ref{eq:J})
for quantum states,
classical discord~(\ref{eq:classicaldiscorddef}) is zero in the ideal case.
However,
discord is nonzero if,
analogous to the quantum case,
which optimizes over all possible measurements,
classical discord also involves noisy measurements.

Alice's measurements are treated} as ideal whereas Bob's measurements are treated as being noisy,
described by a stochastic or a doubly stochastic mapping~$\mathcal{E}$ acting on the state~(\ref{eq:pAB-1MpAB})~\cite{GdOS15}.
The resultant state,
after Bob's noisy measurement,
is
$\bm{p}^{\text{AB}}\mathcal{E}$~(\ref{eq:pAB-1MpAB}).
Whereas the mutual information $I$~(\ref{eq:I}) is known,
the alternative mutual information~(\ref{eq:J})
is modified to include the effect of Bob's noisy measurements and is consequently described by
\begin{equation}
\label{eq:JM}
    J_{\mathcal{E}}^{\text{A};\text{B}}
            \left(\bm{p}^\text{AB}\right)
        :=J^\text{A;B}
            \left(\bm{p}^{\text{AB}}\mathcal{E}\right)
\end{equation}
with the subscript~$\mathcal E$
referring to Bob's noisy channel as we always treat Alice's as ideal: $\mathbb{I}$.
In other words,
the conditional information inherent in inferring alternative mutual information~(\ref{eq:JM})
involves Bob announcing his results to Alice,
and Bob's measurement apparatus is noisy:
described by stochastic or doubly stochastic channels,
as described in~\S\ref{sec:stochasticmap},
prior to ideal measurement and announcement by Bob.
Following this definition of alternative mutual information involving noisy measurement~(\ref{eq:JM}), 
discord for stochastic information is~\cite{GdOS15}
\begin{equation}
\label{eq:classicaldiscorddef'}
\Delta_{\mathcal E}^{\text{A};\text{B}}
    \left(\bm{p}^\text{AB}\right)
    :=I^{\text{A;B}}\left(\bm{p}^\text{AB}\right)
        -J_{\mathcal E}^{\text{A;B}}\left(\bm{p}^\text{AB}\right)
\end{equation}
for specified noisy channel~$\mathcal{E}$,
which only affects~$J$ and not~$I$.
State discord~(\ref{eq:classicaldiscorddef'})
is considered to be classical because states are distributions
and because channels are stochastic maps;
i.e., all the mathematical objects are distributions and their mappings and hence does not require a Hilbert-space description.
This $\mathcal{E}$-dependent discord
is necessarily nonnegative due to the data-processing inequality~\cite{cover2012elements}.

By analogy with quantum discord,
which minimizes over all measurement, classical 
discord corresponds to minimizing state discord~(\ref{eq:classicaldiscorddef'}) over all allowed channels $\{\mathcal{E}\}$~\cite{GdOS15}.
For shared stochastic-information states, classical discord quantifies how much stochasticity is added by a noisy measurement process.
If this noise is described by a doubly stochastic channel,
this noise corresponds to random permutations,
following Birkhoff's theorem,
corresponding to instances of measuring some messages incorrectly as other messages,
with the identity permutation corresponding to measuring all integers correctly. 
Non-zero discord can be interpreted as quantifying stochasticity added by measurement only if entanglement is zero;
otherwise a quantum model is required to describe correlations~\cite{GdOS15}.
Analogous to quantum state merging operationalizing quantum discord~\cite{Horodecki2007},
stochastic-information state merging operationalizes classical discord~\cite{GdOS15}.

Channel distortion,
which is used in rate-distortion theory~\cite{cover2012elements},
quantifies the minimum number of bits required per symbol
that could be achieved over a channel 
so that the input signal can be approximately reconstructed at the output without exceeding a given expected distortion.
Mathematically,
rate-distortion theory,
distortion functions quantify the cost of representing a symbol by an approximate symbol.
Typical distortion functions include Hamming distortion, squared-error distortion and total-variation. 


Total-variation, 
or Kolmogorov,
distance between probability distributions~$\bm{p}$
and $\bm{p}'$~(\ref{eq:pmM}),
namely~\cite{LevinDA2009},
\begin{equation}
\label{eq:tvdistance}
\mathcal{D}\left(\bm{p},\bm{p}'\right)
    :=\nicefrac12\sum_{m\in[M]}
    \left|p_m-p'_m\right|.
\end{equation}
has been widely used for extremum problems,
such as controlling uncertain stochastic systems~\cite{rezaei2012optimal},
approximating a family of probability distributions by a given probability distribution,
maximizing or minimizing entropy subject to total-variation distance constraints,
quantifying uncertainty of probability distributions
by total-variation distance metric, stochastic minimax control, and
in many problems of information, decision theory, and minimax
theory~\cite{charalambous2014extremum}, testing for scale families~\cite{gulati2006Testing} and distortion of channels~\cite{cover2012elements}.  Thus, total-variation distance is well studied and valuable across a broad spectrum of applications,
including for us in comparing total-variation distance to discord.
\section{Approach}
\label{sec:approach}
In this section,
we begin by explaining our model,
which involves three agents:
Alice and Bob who share messages
and Charlie who provides random messages from a distribution.
After describing the model, 
we develop the mathematics required to analyse the effect of noisy measurement in terms of average discord and average distortion in~\S\ref{subsec:mathematics}.
Finally we elaborate on our methods for solving the expressions and what we plot in~\S\ref{subsec:methods}.
\subsection{Model}
\label{subsec:model}
We describe our model for discord as a three-agent protocol involving Charlie,
Alice
and Bob.
By describing the tasks performed by each of the three agents,
we have fully described the protocol
and the pertinent quantifiers of discord and distortion.
Although this model is implied in a previous study of classical discord,
we need to make explicit the agents of this protocol and their actions to be clear in our study of channel discord.

Charlie generates joint distributions
\begin{equation}
\label{eq:pABdist}
    \{\bm{p}^\text{AB}\in\text{mat}_{M^\text{A}\times M^\text{B}}\left(\mathbb{R}^{\geq0}\right)\}
\end{equation}
with prior~$Q\left(\bm{p}^\text{AB}\right)$
and then computes $I^\text{A;B}\left(\bm{p}^\text{AB}\right)$~(\ref{eq:I})
for each~$\bm{p}^\text{AB}$.
For given~$\bm{p}^\text{AB}$~(\ref{eq:pAB}),
Charlie generates a length-$\varsigma$ sequence of pairs of integers
\begin{equation}
\label{eq:lengthvarsigmaseq}
\left\{\left(m^\text{A}\in[M^\text{A}],
    m^\text{B}\in[M^\text{B}]\right)\right\}
\end{equation}
by sampling over~$\bm{p}^\text{AB}$.
In each instance,
the first integer message~$m^\text{A}$ is sent to Alice and the second integer message sent to Bob.
As~A and~B can experience noise in their readout,
the resultant messages,
$m^{\text{A}'}$ and~$m^{\text{B}'}$,
can differ from the original messages,
$m^{\text{A}}$ and~$m^{\text{B}}$.

Alice and Bob send this noisy pair,
$m^{\text{A}'}$ and~$m^{\text{B}'}$,
to Charlie.
At the end of this part of the protocol,
Charlie has stored the length-$\varsigma$ sequence~$\{m^{\text{A}'},m^{\text{B}'}\}$.
Charlie then infers the distribution~$\bm{p}^{\text{A}'\text{B}'}$ from these data,
with this inferred state represented by~$\tilde{\bm{p}}^{\text{A}'\text{B}'}$.

As Alice's instrument is assumed to be noiseless,
$m^{\text{A}'}\equiv m^\text{A}$,
Charlie's procedure is greatly simplified:
he does not send Alice the message but just stores it.
The message pair is thus $\{(m^\text{A},m^{\text{B}'})\}$.
Charlie's inferred state is~$\widetilde{\bm{p}^{\text{AB}}\mathcal{E}}\in\text{mat}_{M^\text{A}\times M^\text{B}}\left(\mathbb{R}^{\geq0}\right)$,
which approximates the actual state~$\bm{p}^{\text{AB}}\mathcal{E}$
after Bob's noisy measurement and~Eq.~(\ref{eq:pAB}).
Charlie computes all permutations of the state~$\widetilde{\bm{p}^{\text{AB}}\mathcal{E}}$,
with each permuted state denoted
$\widetilde{\bm{p}^{\text{AB}}\mathcal{E}}\Pi_\sigma$.

He thence estimates the alternative mutual information $J_\sigma^\text{A;B}$~(\ref{eq:J})
with the subscript~$\sigma$ indicating which of the permuted states~$\widetilde{\bm{p}^{\text{AB}}\mathcal{E}}\Pi_\sigma$ is being considered.
With these results at hand,
Charlie estimates the discord for each
$\widetilde{\bm{p}^{\text{AB}}\mathcal{E}}\Pi_\sigma$
and computes the minimum discord over all~$\sigma$.
Then he averages over results for many generated states~$\bm{p}^\text{AB}$
to obtain an estimate for average discord for a specific channel corresponding to Bob's noisy measurement.

For each estimate
$\widetilde{\bm{p}^{\text{AB}}\mathcal{E}}\Pi_\sigma$,
he computes the distortion,
which he quantifies by the total-variation distance~$\mathcal{D}_\sigma^{\text{A;B}}\left(\widetilde{\bm{p}^{\text{AB}}\mathcal{E}}\Pi_\sigma\right)$, between~$\bm{p}^\text{AB}$ and the estimate
$\widetilde{\bm{p}^{\text{AB}}\mathcal{E}}\Pi_\sigma$. 
Charlie repeats this task for all permutations~$\sigma$
to obtain the minimum total-variation distance
and then averages over all states to obtain
average distortion.
The mathematical description of this procedure is in~\S\ref{subsubsec:estimatingdiscorddistortion}.

In each instance Alice receives noiseless message~$m^\text{A}$,
which is the versor $\bm{\delta}_{m^\text{A}}$,
which she reads
and sends the same message back to Charlie.
Thus, for our mathematical analysis,
Alice's role is superfluous, hence neglected in our protocol.

In each instance Bob receives message~$m^\text{B}$.
His measurement is noisy,
which we describe by a doubly  stochastic channel~$\mathcal{E}$ described in~\S\ref{subsec:statediscord}.
This noise corresponds to permutations of the message basis so some messages are seen to be different messages incorrectly except in the case of the identity permutation~$\mathds1$,
which corresponds to reading  the message correctly.

To elucidate our model, we consider the specific case of a two-bit channel.
Thus, we assume that Charlie generates a single bit each for Alice and Bob;
i.e., $M^A = M^B = 2$ and $m^A, m^B \in \{0, 1\}$.
Let the noisy channel be the mapping of 0 to 0,
i.e., $0\mapsto0$,
with probability $\nicefrac23$.
Then, by the doubly stochastic property,
$0\mapsto1$ with probability~$\nicefrac13$ and $1\mapsto1$ with probability $\nicefrac23$ so $1\mapsto0$ with probability~$\nicefrac13$
the matrix describing this mapping is doubly stochastic.
Bob sends the message~$m^{\text{B}'}$ obtained from his measurement back to Charlie.
Note that,
for the execution of this protocol,
the same stochastic matrix,
which describes Bob's measurement noise,
is applied once per instance,
i.e., each time that Bob reports each measurement outcome.

\subsection{Mathematics}
\label{subsec:mathematics}
In this subsection,
we describe in~\S\ref{subsubsec:generatingnewjointdist} how Charlie generates states as randomly chosen joint distributions to be sent to Alice and Bob,
and then we describe how random channels are generated in~\S\ref{subsubsec:generatingchannels}
for Bob.
In~\S\ref{subsubsec:applyingchannel}
we describe mathematically how the channel is applied to the joint state.
Permutations are applied to states,
and  both discord and distortion are minimimized over all permutations,
as described in~\S\ref{subsubsec:estimatingdiscorddistortion}.
Finally,
in~\S\ref{subsubsec:estimatingdiscorddistortion},
we explain how average discord and average distortion are estimated.
\subsubsection{Generating joint distributions}
\label{subsubsec:generatingnewjointdist}
In this subsubsection,
we explain mathematically how Charlie generates~$\bm{p}^\text{AB}$.
Charlie constructs a prior~$Q\left(\bm{p}^\text{AB}\right)$,
which is heavily weighted over high-entropy states,
meaning that state entropy~(\ref{eq:H}) is close to the upper bound.
By sampling this prior,
Charlie obtains states with high, low and medium entropy
by a linear interpolation between states drawn randomly from~$Q\left(\bm{p}^\text{AB}\right)$
and the state represented by the identity matrix denoted~$\mathds1$.
This linear interpolation generates a continuum of interpolated states
for each of the~$N_\text{rand}$ states.
Thus, Charlie generates
random states from which he draws messages to send to Alice and Bob.

To sample from~$Q\left(\bm{p}^\text{AB}\right)$,
we generate a random $\bm{p}\in\text{mat}_{M^\text{A}\times M^\text{B}}(\mathbb{R})$
with each of the $M=M^\text{A}M^\text{B}$ entries chosen uniformly from~$[0,1]$.
The matrix~$\bm{p}$ is then normalized by dividing all entries by their sum.
This method of selecting random states leads to states that are mostly high-entropy
with the maximum entropy being~$\log\left(M^\text{A}M^\text{B}\right)$.
For $M^\text{A}=M^\text{B}$,
which corresponds to a square matrix,
the maximum entropy is~$\log M$ for~$M$ the square~(\ref{eq:MAMBM}) of~$M^\text{A}$.

Alternatively,
Charlie can sample from a joint-state prior~$Q\left(\bm{p}^\text{AB}_\text{cp}\right)$
for conditionally pure (cp) states,
with conditionally pure states satisfying  Fact~\ref{fact:condpure}.
Conditionally pure states are permutationally equivalent to diagonal matrices
$\text{diag}_{\min\{M^\text{A},M^\text{B}\}}([0,1])$,
as explained in Eq.~(\ref{eq:cpdiagonal}).
Thus, 
For $M^\text{A}=M^\text{B}$,
a conditionally pure state is constructed by generating a random
$\bm{p}_\text{cp}\in\text{diag}_{M^\text{B}}([0,1])$
with each of the diagonal entries chosen uniformly from~$[0,1]$
and then normalized such that the sum of diagonal elements is one.
As for general states,
such conditionally pure states also tend to have high entropies.

Sampling either~$Q\left(\bm{p}^\text{AB}\right)$ or~$Q\left(\bm{p}^\text{AB}_\text{cp}\right)$
yields a candidate state~$\bm{p}_\text{cp}^\text{AB}$,
which is then mapped to a family
according to
\begin{equation}
\label{eq:pfamily}
    \bm{p}_\text{cp}^\text{AB}
        \mapsto
            \frac{a\bm{p}_\text{cp}^\text{AB}+b\mathds1}{\norm{a\bm{p}_\text{cp}^\text{AB}+b\mathds1}}\;
    \forall\,a\in[0,1],\,
    b\in[B]
\end{equation}
where~$B\gg1$ to ensure sufficiently many medium- and low-entropy states.
The channel representing Bob's noisy measurement then acts on Bob's message share, 
and we explain how to generate these channels in the next subsection.
\subsubsection{Generating channels}
\label{subsubsec:generatingchannels}
In this subsubsection, we explain how to generate a random doubly stochastic channel for Bob.
As the doubly stochastic channel is a permutahedron
discussed in~\S\ref{sec:stochasticmap},
generating a random channel is equivalent to constructing the length~$M^\text{B}!$ weight, or probability,
vector~$\bm\wp$,
which corresponds to the probability coefficients for reverse lexicographic ordering of permutations~$\{\sigma\}$,
similarly to reverse lexicographical ordering of the vector of permutation matrices~(\ref{eq:Pivector}).
A given weight vector~$\bm\wp$
has associated entropy
\begin{equation}
\label{eq:weightvectorentropy}
    0\leq H(\bm\wp)\leq\log M^\text{B}!,
\end{equation}
which is the entropy of the corresponding channel~$\mathcal{E}$.
Now we explain how to generate~$\bm\wp$
from a distribution~$\mathcal P$ that is an equal weighting of a uniform prior~$\mathcal{P}_\uparrow$,
resulting in high-entropy weight vectors~$\bm{\wp}_\uparrow$
such that
its entry $H(\bm{\wp}_\uparrow)$~(\ref{eq:H})
is  the maximum entropy~$\log M^\text{B}!$,
and another prior~$\mathcal{P}_\downarrow$
that generates states~$\bm{\wp}_\downarrow$ with low entropy
$H(\bm{\wp}_\downarrow)$~(\ref{eq:H}).

To generate~$\mathcal{P}_\uparrow$,
we first set each entry of~$\bm{\wp}_\uparrow$
be $1$ 
and then normalize this length~$M^\text{B}!$ weight vector
by dividing each element by~$\|\bm\wp\|_1$.
In contrast,
we generate~$\mathcal{P}_\downarrow$
by uniformly randomly generating the first element of $\bm\wp_\downarrow$,
namely~$\wp_\downarrow^1$,
from the interval $[0,1]$
and replace
\begin{equation}
\wp_\downarrow^1
\gets\frac{\wp_\downarrow^1}{M^\text{B}!}.
\end{equation}
The next element of the weight vector,
namely, $\wp_\downarrow^2$,
is drawn uniformly from the interval $$\left[1-\wp_\downarrow^1, 1\right],$$
and we continue according to the rule that $\wp_\ell$ is drawn uniformly from $$\left[1-\sum_{\jmath=1}^{\ell-1}\wp_\downarrow^\jmath,1\right].$$
After randomly generating all these~$M^\text{B}!$ elements,
we normalise this weight vector to obtain  $\bm{\wp}_{\downarrow}$.

Now that we have generated an instance of a low-entropy weight vector~$\bm{\wp}_\downarrow$
and a high-entropy weight vector~$\bm{\wp}_\uparrow$
we generate numerous vectors in between the two by linear interpolation,
thereby sampling the continuous set
\begin{equation}
\label{eq:weightlincomb}
    \bm\wp\left(\bm{\wp}_\downarrow,\bm{\wp}_\uparrow,a\right)
        :=\frac{a\bm{\wp}_\downarrow+(1-a)\bm{\wp}_\uparrow}{\norm{a\bm{\wp}_\downarrow+(1-a)\bm{\wp}_\uparrow}}\,\forall a\in[0,1],
\end{equation}
This interpolation~(\ref{eq:weightlincomb}) yields medium-entropy channels to round out the sampling.

As we have generated many random instances of $\bm\wp$~(\ref{eq:weightlincomb}),
we can construct corresponding descriptions of doubly stochastic channels.
The elements of~$\bm\wp$
are coefficients of reverse  lexicographically ordered permutation matrices,
and this weighted sum is then the permutahedron that describes random doubly stochastic channel~$\mathcal{E}$ with entropy given by the entropy of its representative weight vector.
Mathematically,
the matrix description of the channel is
\begin{equation}
\label{eq:channeldescription}
    \mathcal{E}
        =\bm{\wp}\cdot\bm{\Pi},\;
        \bm{\Pi}:=\left(\Pi_\sigma;\sigma\in S_{M^\text{B}!}\right),
\end{equation}
for~$\bm{\wp}$ and~$\bm{\Pi}$ length-$M^\text{B}!$ vectors of real numbers and permutation matrices~(\ref{eq:Pivector}), respectively,
and~$\sigma\in S_{M^\text{B}!}$
is drawn
in reverse lexicographical order.
\subsubsection{Applying the channel to the joint distributions}
\label{subsubsec:applyingchannel}
To describe Bob's noisy measurement mathematically,
we apply the generated random channel, 
discussed in~\S\ref{subsubsec:generatingchannels},
to Bob's share of the entire message~$\bm{p}^\text{AB}$
sent by Charlie.
First suppose that Alice and Bob each have noisy measurements described by channels~$\mathcal{E}^\text{A}$
and~$\mathcal{E}^\text{B}$,
respectively.
Then the state sent back to Charlie from Alice and Bob is
\begin{equation}
\label{eq:1MpAB}
    \mathcal{E}^\text{A}\bm{p}^\text{AB}\mathcal{E}^\text{B}\in\text{mat}_{M^\text{A}\times M^\text{B}}(\mathbb{C}),
    \bm{p}^\text{AB}\in\text{mat}_{M^\text{A}\times M^\text{B}}(\mathbb{C}),
    \mathcal{E}^\text{A}\in\text{mat}_{M^\text{A}\times M^\text{A}}(\mathbb{C}),
    \mathcal{E}^\text{B}\in\text{mat}_{M^\text{B}\times M^\text{B}}(\mathbb{C}),
\end{equation}
corresponding to one application of a noisy channel for each of Alice's and Bob's noisy measurements but noiseless transmission back to Charlie.
As Alice's measurement is noiseless,
we assign
\begin{equation}
\mathcal{E}^\text{A}\equiv\mathds1,\;
\mathcal{E}^\text{B}=:\mathcal{E}.
\end{equation}
\begin{remark}
In~\S\ref{subsec:model},
Charlie estimates~$\bm{p}^{\text{AB}}\mathcal{E}$
to be~$\widetilde{\bm{p}^{\text{AB}}\mathcal{E}}$  
by repeated sampling but here,
in the mathematical description,
we work with exact descriptions of the state~(\ref{eq:1MpAB}).
\end{remark}
\subsubsection{Estimating discord and distortion}
\label{subsubsec:estimatingdiscorddistortion}
After Charlie receives
$\bm{p}^{\text{AB}}\mathcal{E}$~(\ref{eq:1MpAB}),
he computes all permutations of this matrix.
Charlie generates each of the $M^\text{B}!$
instances of $M^\text{B}\times M^\text{B}$ permutation matrices $\Pi_\sigma$~(\ref{eq:Pisigmap}), 
for each~$\sigma$ drawn from permutation group $S_{M^\text{B}}$
in  reverse lexicographical order.
For each instance~$\sigma$,
Charlie obtains the permuted state by multiplying
$\bm{p}^\text{AB}\mathcal{E}$ by each permutation matrix~$\Pi_\sigma$.

Here we redefine average discord
to quantify channel distortion,
in contrast to the earlier definition of classical discord in terms of fluctuations for a state~\cite{GdOS15},
and we define average distortion.
This redefinition of average discord,
in terms of channels,
then allows us to show numerically the monotonic relationship between classical (channel) discord and channel distortion.
Average discord is obtained first by minimizing discord~(\ref{eq:classicaldiscorddef'})
over all state permutations and then by averaging over all states according to the prior~$Q\left(\bm{p}^\text{AB}\right)$ in~\S\ref{subsubsec:generatingnewjointdist}.
Similarly,
average distortion is obtained
by averaging state-dependent distortion~(\ref{eq:tvdistance}) 
over the prior of states~$Q\left(\bm{p}^\text{AB}\right)$.

For average discord,
we first extend state-dependent discord~(\ref{eq:classicaldiscorddef'})
to be the minimized state-dependent discord over all permutations of the state
\begin{equation}
\label{eq:averagediscord'}
\text{min}_\sigma
    \Delta^{\text{A};\text{B}}_{\mathcal{E}}
    \left(\bm{p}^\text{AB}\Pi_\sigma\right).
\end{equation}
Averaging over the prior~$Q\left(\bm{p}^\text{AB}\right)$
yields channel discord
\begin{equation}
\label{eq:averagediscordintegral}
    \Delta^{\text{A};\text{B}}\left(\mathcal{E}\right)
        :=\int\text{d}Q\left(\bm{p}^\text{AB}\right)
    \text{min}_\sigma
    \Delta^{\text{A};\text{B}}_{\mathcal{E}}
    \left(\bm{p}^\text{AB}\Pi_\sigma\right),
\end{equation}
which quantifies the discord due to noisy measurement in a state-independent but of course prior-dependent way.
Average discord is obtained by sampling the integral~(\ref{eq:averagediscordintegral})
to obtain the estimate~$\tilde{\Delta}^{\text{A};\text{B}}(\mathcal{E})$.

Similarly, we extend the definition of distortion $\mathcal{D}$~(\ref{eq:tvdistance}) first by minimizing over permutations and then by averaging over the prior~$Q\left(\bm{p}^\text{AB}\right)$.
The state- and channel-dependent distortion is
\begin{equation}
\label{eq:minavdiscord}
    \mathcal{D}^{\text{A;B}}_{\mathcal E}
    \left(\bm{p}^{\text{AB}}\right)
    :=\mathcal{D}\left(\bm{p}^{\text{AB}}\mathcal{E},\bm{p}^\text{AB}\right)
\end{equation}
and its minimization over all permutations is
\begin{equation}
\label{eq:averagedistortion}
   \text{min}_\sigma\mathcal{D}_{\mathcal{E}}^{\text{A;B}}\left(\bm{p}^\text{AB}\Pi_\sigma\right),
\end{equation}
Then,
analogous to average discord~(\ref{eq:averagediscordintegral}),
we integrate to obtain channel distortion
\begin{equation}
\label{eq:averagedistortionintegral}
\mathcal{D}^{\text{A};\text{B}}\left(\mathcal{E}\right)
    :=\int\text{d}Q\left(\bm{p}^\text{AB}\right)\text{min}_\sigma\mathcal{D}_{\mathcal{E}}^{\text{A;B}}\left(\bm{p}^\text{AB}\Pi_\sigma\right)
\end{equation}
for a given channel.
Average discord is obtained by sampling the integral~(\ref{eq:averagediscordintegral})
to obtain the estimate~$\tilde{\mathcal D}^{\text{A};\text{B}}(\mathcal{E})$.

\subsubsection{Example: Two-bit channel}
\label{subsubsec:mathematics/twobit}
We now elucidate our model by considering the special case where
Charlie generates a single bit for Alice and a single bit for Bob:
$M^A = M^B = 2$.
Thus,
\cref{eq:pAB} implies that the joint probability of messages
    shared between Alice and Bob is specified by the matrix
    \begin{equation}
    \label{eq:twobitstate}
      \bm{p}^\text{AB} = 
      \begin{pmatrix}
        p_{00} & p_{01} \\
        p_{10} & p_{11}
      \end{pmatrix},\ p_{ij} \geq 0,\ 
      p_{00} + p_{01} + p_{10} + p_{11} = 1.
\end{equation}
The two-bit state~(\ref{eq:twobitstate})
is subjected to a distortion channel whose form is dictated by \cref{eq:channeldescription},
which implies
\begin{equation}
    \label{eq:twobitchannel}
      \mathcal{E} =
      \begin{pmatrix} 1 - \mu \\ \mu \end{pmatrix} \cdot
      \begin{pmatrix} \mathds{1} \\ X \end{pmatrix} =
      (1 - \mu) \mathds{1} + \mu X =
      \begin{pmatrix}
          1 - \mu & \mu \\ \mu & 1 - \mu
      \end{pmatrix},\ 0 \leq \mu \leq 1.
    \end{equation}
Consequently, the entropy is
\begin{equation}
\label{eq:twobitentropy}
H = -(1-\mu) \log (1-\mu) - \mu \log\mu
\end{equation}
according to \cref{eq:weightvectorentropy}.

Applying the channel to the two-bit state yields
    \begin{equation}
       \bm{p}^\text{AB} \mathcal{E} =
       \begin{pmatrix}
        p_{00} & p_{01} \\
        p_{10} & p_{11}
      \end{pmatrix} \cdot
      \begin{pmatrix}
          1 - \mu & \mu \\ \mu & 1 - \mu
      \end{pmatrix} =
        \begin{pmatrix}
         (1-\mu) p _{00}+\mu  p _{01} & \mu  p _{00}+(1-\mu) p _{01} \\
         (1-\mu) p _{10}+\mu  p _{11} & \mu  p _{10}+(1-\mu) p _{11} \\
        \end{pmatrix}.
    \end{equation} 
We then compute the state discord per \cref{eq:averagediscord'},
and we could perform a similar calculation for state distortion per \cref{eq:tvdistance}.
The explicit expression for discord~(\ref{eq:averagediscord'}) is
\begin{align}
\label{eq:twobitdiscordmu}
   & \left( (1-\mu) p_{00} + \mu p_{01} + (1-\mu) p_{10} + \mu p_{11} \right)
   \log \left( (1-\mu) p_{00} + \mu p_{01} + (1-\mu) p_{10} + \mu p_{11} \right) \nonumber \\
 + & \left( \mu p_{00} + (1-\mu) p_{01} + \mu p_{10} + (1-\mu) p_{11} \right)
   \log \left( \mu p_{00} + (1-\mu) p_{01} + \mu p_{10} + (1-\mu) p_{11} \right) \nonumber \\
 - & \left( (1-\mu) p_{00} + \mu p_{01} \right)
     \log \left( (1-\mu) p_{00} + \mu p_{01} \right)
 -   \left(\mu  p_{00} + (1-\mu) p_{01} \right)
     \log \left(\mu  p_{00} + (1-\mu) p_{01} \right) \nonumber \\
 - & \left( (1-\mu) p_{10} + \mu p_{11} \right) \log \left( (1-\mu) p_{10} + \mu p_{11} \right)
 -   \left( \mu p_{10} + (1-\mu) p_{11} \right) \log \left( \mu p_{10} + (1-\mu) p_{11} \right) \nonumber \\
 - & \left( p_{00} + p_{10} \right) \log \left( p_{00} + p_{10} \right)
 -   \left( p_{01} + p_{11} \right) \log \left( p_{01} + p_{11} \right) \nonumber \\
 + & p_{00} \log \left( p_{00} \right) + p_{01} \log \left( p_{01} \right)
 +   p_{10} \log \left( p_{10} \right) + p_{11} \log \left( p_{11} \right)
\end{align}
as a function of state~$\bm{p}^\text{AB}$ and channel parameter $\mu$.

If we could integrate discord and distortion over all possible joint states, i.e.,
over the formal variables
$p_{00}, p_{01}, p_{10}, p_{l1}$ with respect to an appropriate prior
distribution, then we could obtain channel discord and distortion via
\cref{eq:averagediscordintegral,eq:averagedistortionintegral},
respectively. Note that the resulting expressions depend only on $\mu$.
On the other hand, if discord is monotonic with respect to entropy for all states~$\{\bm{p}^\text{AB}\}$,
i.e., for all points on the tetrahedron,
then discord would be monotonic with respect to this average.

\subsection{Methods}
\label{subsec:methods}
In this subsection we discuss how we average discord and average distortion for many randomly chosen channels,
with each of these channels corresponding to a different noisy measurement process implemented by Bob.
First,
we explain in~\S\ref{subsubsec:numericallygenstates} how many instances of states we generate and how to create those instances.
Then,
in~\S\ref{subsubsec:numericallygenpermu},
we explain how we generate all permutation matrices and thence all permuted states.
Then we explain how we generate random channels~$\{\mathcal{E}\}$ numerically in~\S\ref{subsubsec:numericallygenrandchannels}
and how many such channels. 
Finally,
in~\S\ref{subsubsec:relatingavdiscavdist} we explain how we calculate the states sent back to Charlie, 
how to use these states to compute average discord and average distortion, and then study their relations.
\subsubsection{Numerically generating states}
\label{subsubsec:numericallygenstates}
We begin by generating joint states 
from either prior~$Q\left(\bm{p}^\text{AB}\right)$
or~$Q\left(\bm{p}^\text{AB}_\text{cp}\right)$
as described in~\S\ref{subsubsec:generatingnewjointdist}.
In practice,
we generate random states as follow.
First we choose $B=99$ in Eq.~(\ref{eq:pfamily})
as we have discovered empirically that this value of~$B$ yields a good spread of low-, medium- and high-entropy states.
Then we step through values of linear interpolation parameter~$a$
by increasing in step sizes that grow quadratically:
the coefficient $a$ of~$\mathds1$
increases in steps of $([\varpi-1]0.0101)^2$
for $\varpi\in[100]$.

For each choice of~$a$ and fixing~$b=(1-a)B$,
we choose a new random instance of~$\bm{p}^\text{AB}$
according to the random-matrix construction method described in~\S\ref{subsubsec:generatingnewjointdist}.
We insert~$a$, $b$ and~$\bm{p}^\text{AB}$
into Eq.~(\ref{eq:pfamily}) to obtain one instance of a state for calculating average discord and average distortion.
\subsubsection{Numerically generating all permutation matrices}
\label{subsubsec:numericallygenpermu}
Our analysis considers different messages sizes~$M$ such that
$M^\text{A}=M^\text{B}=\sqrt{M}$,
where we have chosen to study only cases for which Alice's and Bob's messages are the same size.
For each message size 
all permutation matrices are generated according to the mathematical description in~\S\ref{sec:probvec}.
We need to construct~$M^\text{B}!$ permutation matrices,
each of size $M^\text{B}\times M^\text{B}$ ,
representing permutations $\sigma\in S_{M^\text{B}}$.
Each row of the matrix corresponds to a permuted version of the length-$M^\text{B}$ initial row vector,
which is the vector comprising entries that are the column numbers themselves.
These columns correspond to reverse lexicographical ordering of the permutations,
with  reverse lexicographical ordering discussed in~\S\ref{sec:probvec}.

We apply \textsc{perms} from MATLAB$^{\textregistered}$
to the initial vector,
and the output is the $M^\text{B}!\times M^\text{B}$
matrix~$\Theta$,
whose elements are column indices for nonzero entries of permutation matrices~$\Pi_\sigma$.
For each of the~$M^\text{B}!$ rows,
labelled by~$\sigma$ in reverse lexicographical order,
we construct the corresponding
$M^\text{B}\times M^\text{B}$
permutation matrix~$\Pi_\sigma$,
whose entries are all zeroes and ones such that only one instance of one appears in each row or column as described in
and around Eq.~(\ref{eq:Pisigmap}).

Specifically, to generate~$\Pi_\sigma$,
we pick row~$\sigma$ from the matrix~$\Theta$,
denoted as a vector~$\bm{\Theta}_\sigma$. 
The value of the entry in the first column of~$\bm{\Theta}_\sigma$ indicates which element of the first row of~$\Pi_\sigma$ is one,
with the rest of the elements in that row being zero.
Then we proceed to the second entry of the vector~$\bm{\Theta}_\sigma$,
and its value indicates which element of the second row of~$\Pi_\sigma$ is one. 
We continue for all~$M^\text{B}$ rows of~$\Pi_\sigma$
and then repeat for all $\sigma\in S_{M^\text{B}}$.
In this way we have constructed the full set of permutation matrices for message~$m^\text{B}$.

\subsubsection{Numerically generating random channels}
\label{subsubsec:numericallygenrandchannels}
We generate 6000 doubly stochastic channels,
each channel represented by some weight vector
$\bm{\wp}\left(\bm{\wp}_\downarrow,a\right)$~(\ref{eq:weightlincomb}),
regardless of message size.
Construction of each of these weight vectors proceeds according to the mathematical description in~\S\ref{subsubsec:generatingchannels}.
We choose to generate 6000 instances of random channels because we allow for 100 equally spaced values of~$a$ in Eq.~(\ref{eq:weightlincomb})
and 60 randomly chosen~$\bm{\wp}_\downarrow$
for each~$a$.

A doubly stochastic channel is a permutahedron,
and we have generated the set of all permutation matrices~$\Pi_\sigma\in S_{M^\text{B}}$
in~\S\ref{subsubsec:numericallygenpermu}.
Specifically,
we generate random weight vectors according to Eq.~(\ref{eq:weightlincomb}),
thereby yielding low-, medium- and high-entropy~(\ref{eq:weightvectorentropy})
weight vectors.
The resultant set of 6000 randomly generated weight vectors faithfully represents 6000 randomly generated channels,
and our interpolation~(\ref{eq:weightlincomb})
ensures good sampling of a wide range of channel entropies. 
In addition, we manually add the noiseless channel~$\mathds1$ to our simulations to
include the instance of zero average discord and zero average distortion.
\subsubsection{Relating average discord to average distortion}
\label{subsubsec:relatingavdiscavdist}
Now that state~$\bm{p}^\text{AB}$ is generated numerically according to the procedure described in~\S\ref{subsubsec:numericallygenstates},
for both~$Q\left(\bm{p}^\text{AB}\right)$
corresponding to random initial states and~$Q\left(\bm{p}^\text{AB}_\text{cp}\right)$ for random conditionally pure states,
we calculate the corresponding~$\bm{p}^{\text{AB}}\mathcal{E}$
for each permutation~$\sigma$.
These permuted returned states~$\{\bm{p}^{\text{AB}}\mathcal{E}\Pi_\sigma\}$ are used to calculate both average discord and average distortion.
Mathematical expressions for average discord and average distortion are given in~\S\ref{subsubsec:estimatingdiscorddistortion}.

We begin with how we calculate average discord 
$\Delta^{\text{A};\text{B}}\left(\mathcal{E}\right)$~(\ref{eq:averagediscordintegral}). 
First we calculate each~$\Delta^{\text{A};\text{B}}_\mathcal{E}\left(\bm{p}^\text{AB}\Pi_\sigma\right)$ 
for each~$\bm{p}^{\text{AB}}$
and then minimize over all~$\sigma\in S_{M^\text{B}}$
according to Eq.~(\ref{eq:averagediscord'}),
thereby obtaining the minimized~$\Delta^{\text{A};\text{B}}\left(\mathcal{E}\right)$.
The next step is to average over all~$\bm{p}^{\text{AB}}$.
As explained in~\S\ref{subsubsec:numericallygenstates},
we generate a random state~$\bm{p}^{\text{AB}}$
for each choice of linear interpolation parameter $a$~(\ref{eq:weightlincomb}),
which suffices to sample the integral~(\ref{eq:averagediscordintegral})
fairly and thus obtain a good estimate~$\tilde{\Delta}^{\text{A};\text{B}}(\mathcal{E})$
of the actual average discord~$\Delta^{\text{A};\text{B}}(\mathcal{E})$.

The procedure for calculating average distortion  $\mathcal{D}^{\text{A};\text{B}}(\mathcal{E})$~(\ref{eq:averagedistortionintegral}) is similar. 
First we calculate each~$\mathcal{D}^{\text{A};\text{B}}_\mathcal{E}\left(\bm{p}^\text{AB}\Pi_\sigma\right)$ 
for each~$\bm{p}^{\text{AB}}$
and then minimize over all~$\sigma\in S_{M^\text{B}}$
according to Eq.~(\ref{eq:averagedistortion}),
thereby obtaining the minimized~$\mathcal{D}^{\text{A};\text{B}}\left(\mathcal{E}\right)$.
The next step is to average over all~$\bm{p}^{\text{AB}}$.
As explained in~\S\ref{subsubsec:numericallygenstates},
we generate a random state~$\bm{p}^{\text{AB}}$
for each choice of linear interpolation parameter $a$~(\ref{eq:weightlincomb}),
which suffices to sample the integral~(\ref{eq:averagedistortionintegral})
fairly and thus obtain a good estimate~$\tilde{\mathcal{D}}^{\text{A};\text{B}}(\mathcal{E})$
of the actual average distortion~$\mathcal{D}^{\text{A};\text{B}}(\mathcal{E})$.

Finally,
we relate estimated average discord to estimated average distortion by plotting~$\tilde{\Delta}^{\text{A};\text{B}}(\mathcal{E})$
against~$\tilde{\mathcal{D}}^{\text{A};\text{B}}(\mathcal{E})$. 
Specifically,
we choose sufficiently large yet tractable message sizes, namely, $M^\text{B}\in\{6,7\}$
and make distinct plots for each~$M^\text{B}$.
For each randomly chosen channel,  the resultant single point on the graph corresponding to~$\tilde{\Delta}^{\text{A};\text{B}}(\mathcal{E})$
and~$\tilde{\mathcal{D}}^{\text{A};\text{B}}(\mathcal{E})$
is marked,
and we thereby obtain a scatter plot.
We create plots for two cases,
random initial states and random initial conditionally pure states,
and compare these two cases.

\subsubsection{Example: Two-bit channel}
\label{subsubsec:methods/twobit}
We explicitly analyze the relationship between average channel discord~(\ref{eq:averagediscordintegral})
and channel distortion~(\ref{eq:averagedistortionintegral})
in the case of a two-bit channel as described in \cref{subsubsec:mathematics/twobit}.
By doing so, we establish the monotonicity of discord
as a function of channel entropy. Our approach follows.
\begin{enumerate}
    \item We first recognize that the set of possible channels 
    $\mathcal{E}$ can be parametrized by a single number
    $0 \leq \mu \leq 1$ per \cref{eq:twobitchannel}.
    Hence, the channel discord can be expressed as a function of $\mu$.
    Furthermore, the channel discord is defined to be an integral with
    respect to some measure over an integrand that itself depends on
    $\mu$. If we prove that the integrand is monotonic in $\mu$ over
    some interval, we will also have proved that the integral is
    monotonic in $\mu$ over the same interval.
    \item We can therefore assess the monotonicity of the channel discord~(\ref{eq:averagediscord'}),
    which we write here simply as~$\Delta$,
    as a function of $H$~(\ref{eq:twobitentropy})
    by examining the derivative of
    $\Delta$ with respect to $H$ and applying the chain rule:
\begin{equation}
\label{eq:dDeltadH}
    \dv{\Delta}{H} = \frac{\dv{\Delta}{\mu}}{\dv{H}{\mu}}
    = \left( \log(1-\mu) - \log\mu \right)^{-1} \cdot \dv{\Delta}{\mu}.
\end{equation}
    This expression is well-defined except when $\mu = \nicefrac12$,
    but this point corresponds to the maximum value of $H$
    and so is not relevant for monotonicity arguments. Thus,
    monotonicity of $\Delta$ with respect to $H$ can be proven
    by demonstrating monotonicity on the intervals
    $0 < \mu < \nicefrac12$ and $\nicefrac12 < \mu < 1$,
    which can be proven by demonstrating monotonicity of the
    integrand as described in the previous point.
\end{enumerate}

The above two points imply that we can prove the monotonicity of
channel discord as a function of channel entropy by showing that
the state discord is a monotonically increasing (decreasing) function
of $\mu$ for $0 < \mu < \nicefrac12$ ($\nicefrac12 < \mu < 1$). We prove that this
is so in~\S\ref{subsec:results/twobit}.

\section{Results}
\label{sec:results}
In this section we present our results,
which are numerical in nature.
We choose tractable message-size values,
namely $M^\text{B}\in\{6,7\}$,
to study the relation between average discord~$\tilde{\Delta}^{\text{A;B}}(\mathcal{E})$ and average distortion~$\tilde{\mathcal D}^{\text{A;B}}(\mathcal{E})$.
Specifically,
we plot~$\tilde{\Delta}^{\text{A;B}}(\mathcal{E})$
vs~$\tilde{\mathcal D}^{\text{A;B}}(\mathcal{E})$
for many generated channels as described in~\S\ref{subsubsec:numericallygenrandchannels}
averaged over randomly generated states.
We plots for two cases:
randomly generated joint distributions
in~\S\ref{subsec:plotsranjointstates}
and randomly generated conditionally pure states in~\S\ref{subsec:plotsrancpstates}.
Then we explain our best-fit quadratic relation between~$\tilde{\Delta}^{\text{A;B}}(\mathcal{E})$
and~$\tilde{\mathcal D}^{\text{A;B}}(\mathcal{E})$
in~\S\ref{subsec:bestfit}.
Our methods for generating these plots are described in~\S\ref{subsec:methods}.
\subsection{Plots for randomly generated joint states}
\label{subsec:plotsranjointstates}
\begin{figure}
\centering
\includegraphics[width=\linewidth]{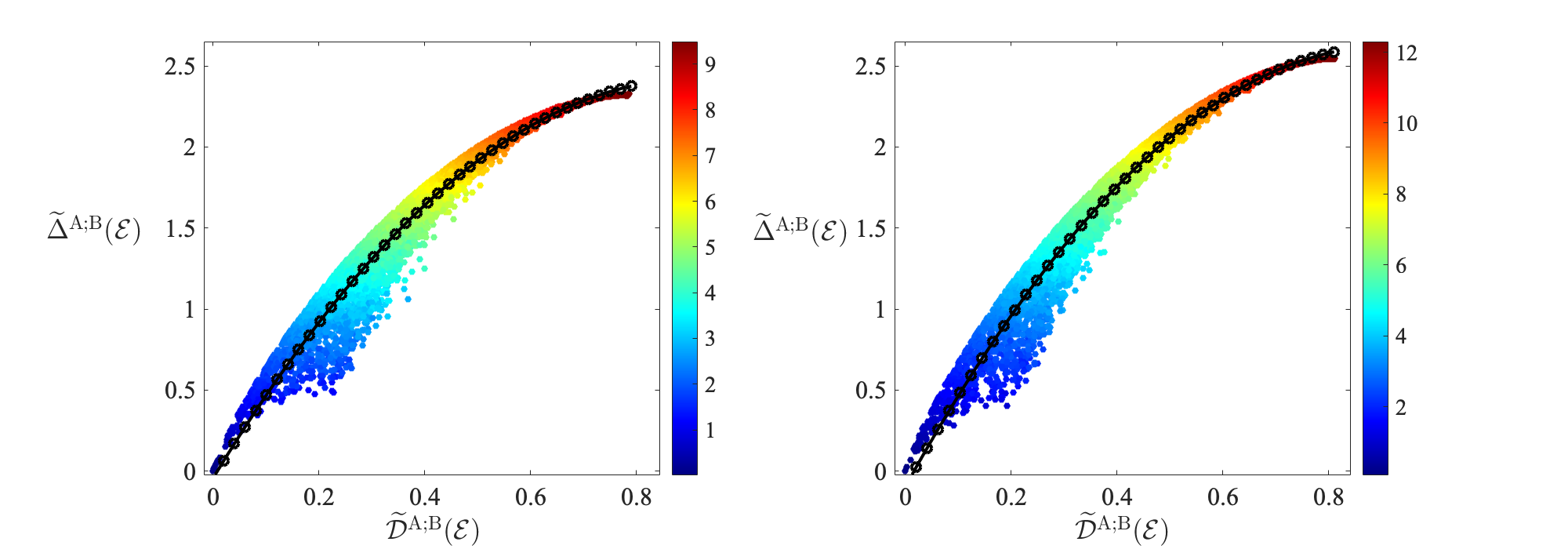}
\caption{%
Scatter plot of average discord~$\tilde{\Delta}^{\text{A;B}}(\mathcal{E})$
vs average distortion~$\tilde{\mathcal D}^{\text{A;B}}(\mathcal{E})$
for Bob's random noisy measurements represented by 6001 randomly chosen doubly stochastic channels $\{\mathcal{E}\}$.
Each instance of~$\mathcal{E}$ is evaluated for 100 random
initial bipartite states $\{\bm{p}^\text{AB}\}$,
described in~\S\ref{subsubsec:numericallygenstates}},
with sizes being
(a) $M^\text{A}=M^\text{B}=6$ and
(b) $M^\text{A}=M^\text{B}=7$.  
The representative weight-vector ($\bm\wp$) entropy for each $\mathcal{E}$
is colour-coded in the heat map ranging from zero to (a)~$\log6!$ and (b)~$\log7!$ 
following the method described in~\S\ref{subsec:plotsranjointstates}.
The black points correspond to a least-squares fit of the quadratic relation~(\ref{eq:quadraticfit}) with
(a) $t_1=-2.952$,
$t_2=5.395$,
$t_3=-0.0446$
with RMSE 0.1005 
and (b) $t_1=-3.223$,
$t_2=5.924$,
$t_3=-0.0971$
with RMSE 0.09854.%

\label{fig:ave_results-bipartite}
\end{figure}

In Fig.~\ref{fig:ave_results-bipartite},
we have plotted estimated average discord $\tilde{\Delta}^{\text{A;B}}(\mathcal{E})$~(\ref{eq:averagediscordintegral}) and 
estimated average distortion $\tilde{\mathcal D}^{\text{A;B}}(\mathcal{E})$~(\ref{eq:averagedistortionintegral})
for two cases of total message length
$M\in\{6,7\}$ as discussed in~\S\ref{subsubsec:relatingavdiscavdist}. 
This scatter plot represents 6001 instances of randomly chosen channels for Bob and randomly chosen initial states by Charlie,
and the points are colour-coded by the Shannon entropy of the weight vector representing the channel~(\ref{eq:weightvectorentropy}).

The origin of the plot corresponds to zero average discord and zero average distortion and arises for Bob's measurement being noiseless, i.e.,
for a zero-entropy weight vector $\bm\wp$.
We observe a monotonic trend of increasing average discord with respect to increasing average distortion.
This monotonicity inference is reinforced in~\S\ref{subsec:bestfit}
where we explain the best-fit curve,
which is certainly monotonic.
Furthermore,
based on the colour-coded heat map in Fig.~\ref{fig:ave_results-bipartite},
we see monotonically increasing of all three: average discord, average distortion and channel entropy.

The scatter plot shows more features.
The highest point of the curve
has the maximumn allowed entropy $\log M^\text{B}!$~(\ref{eq:weightvectorentropy}) for the channel.
For the two chosen messages sizes, the maximum entropies are
\begin{equation}
\label{eq:log67}
    \log6!=9.492,\;
    \log7!=12.299,
\end{equation}
respectively.
Also the scatter plot is narrow for low- and high-entropy cases of channels and wide for medium-low choices of channel entropy.

We have provided Fig.~\ref{fig:ave_results-bipartite}(a)
and Fig.~\ref{fig:ave_results-bipartite}(b)
showing scatter plots for $M^\text{B}=6$
and $M^\text{B}=7$, respectively.
The two scatter plots are similar.
The differences are that the maximum entropy for the second scatter plot is higher due to the larger message size,
with the increase in maximum entropy given by the ratio of the numbers in~(\ref{eq:log67}).
Both estimated average discord and estimated average distortion are increased slightly for increased message size.

\subsection{Plots for randomly generated conditionally pure states}
\label{subsec:plotsrancpstates}
\begin{figure}
\includegraphics[width=\linewidth]{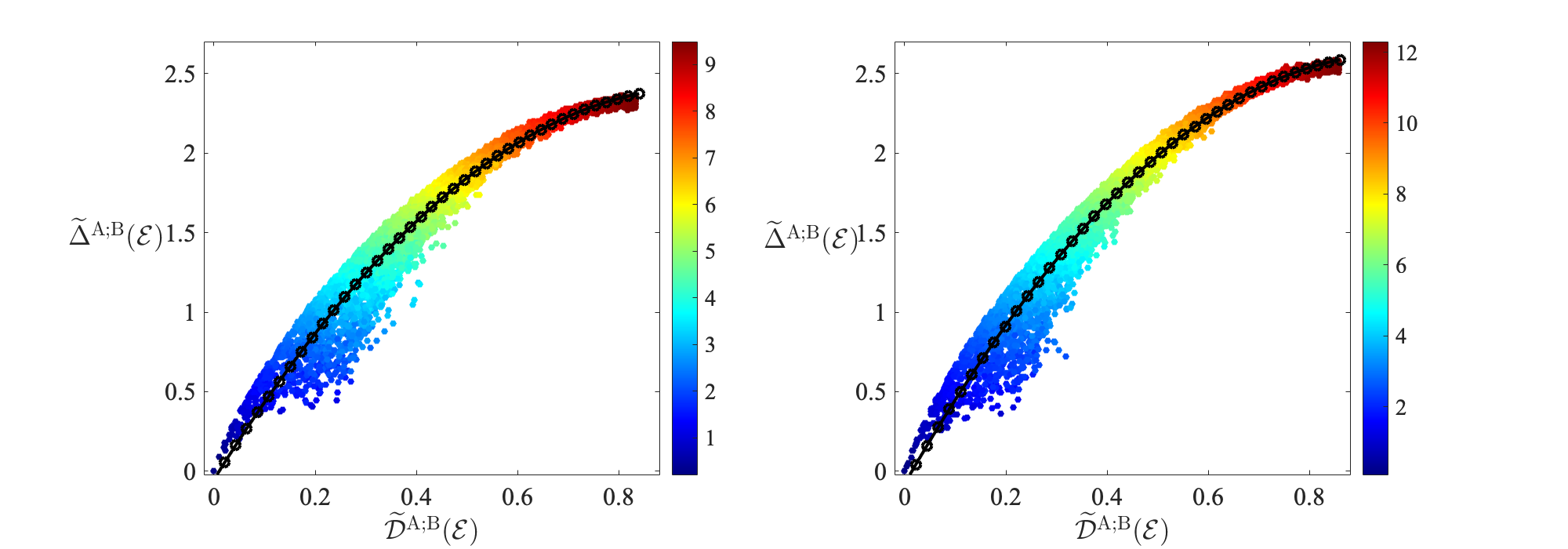}
\caption{%
Scatter plot of average discord~$\tilde{\Delta}^{\text{A;B}}(\mathcal{E})$
vs average distortion~$\tilde{\mathcal D}^{\text{A;B}}(\mathcal{E})$
for Bob's random noisy measurements represented by 6001 randomly chosen doubly stochastic channels $\{\mathcal{E}\}$.
Each instance of~$\mathcal{E}$ is evaluated for 100 random
initial conditionally pure states $\{\bm{p}_\text{cp}^\text{AB}\}$,
described in~\S\ref{subsubsec:numericallygenstates},
with sizes being
(a) $M^\text{A}=M^\text{B}=6$ and
(b) $M^\text{A}=M^\text{B}=7$.  
The representative weight-vector ($\bm\wp$) entropy for each $\mathcal{E}$
is colour-coded in the heat map ranging from zero to (a)~$\log6!$ and (b)~$\log7!$ 
following the method described in~\S\ref{subsec:plotsranjointstates}.
The black points correspond to a least-squares fit of the quadratic relation~(\ref{eq:quadraticfit}) with
(a) $t_1=-2.658$, $t_2=5.119$, $t_3=-0.052$
and RMSE 0.1041
and $t_1=-2.85$, $t_2=5.548$, $t_3=-0.0778$ and RMSE 0.1019.%
}
\label{fig:ave_results_CondiPure}
\end{figure}
In this subsection 
we obtain scatter plots of estimated average discord vs estimated average discord for the case that Charlie generates random conditionally pure states~(\ref{eq:cpdiagonal})
instead of random joint states~(\ref{eq:pAB}) 
as was done in~\S\ref{subsec:plotsranjointstates}.
Other than using conditionally pure states here,
we follow exactly the same procedure used to obtain Fig.~\ref{fig:ave_results-bipartite}.
The purpose of this subsection is to verify or refute that the two cases of initial random joint distributions vs initial random conditionally pure states show the same or different features.

The scatter plot for estimated average discord vs estimated average distortion is shown in Fig.~\ref{fig:ave_results_CondiPure} for initial conditionally pure states.
Similarly to  Fig.~\ref{fig:ave_results-bipartite},
the scatter shows monotonically increasing average discord with respect to average distortion,
monotonicity of both with respect to channel entropy represented by the heat map,
and wider scatter for medium entropy compared to narrow scatter width for low and high entropy.
The differences are only with respect to randomness of generated channels in the plots,
suggesting that Figs.~\ref{fig:ave_results-bipartite}
and~\ref{fig:ave_results_CondiPure}
are identical up to random-sampling variability.
\subsection{Quadratic best fit to the plots}
\label{subsec:bestfit}
Now we explain how we fit a curve to the scatter plots of Figs.~\ref{fig:ave_results-bipartite} and~\ref{fig:ave_results_CondiPure}.
Our numerical results fit well with a quadratic curve
\begin{equation}
\label{eq:quadraticfit}
    \tilde{\Delta}^{\text{A;B}}(\mathcal{E})
        =t_1\left(\tilde{\mathcal D}^{\text{A;B}}(\mathcal{E})\right)^2+t_2\tilde{\mathcal D}^{\text{A;B}}(\mathcal{E})+t_3
\end{equation}
with $\{t_\imath\}$ chosen differently for each plot to minimize root-mean-square error (RMSE).
In all four cases,
RMSE$\sim0.1$,
which indicates a good fit as the RMSE is much smaller than the range of~$\tilde{\Delta}^{\text{A;B}}(\mathcal{E})$.

\subsection{Monotonicity of two-bit channel discord as function of channel entropy}
\label{subsec:results/twobit}
Now we show that the discord of a two-bit noise channel varies 
monotonically with the entropy of that channel.
As we explain in \cref{subsubsec:methods/twobit},
it suffices to show that the \emph{state} discord $\Delta$
for an arbitrary two-bit state~(\ref{eq:twobitstate})
varies monotonically in the parameter 
$\mu$ that specifies the channel~(\ref{eq:twobitchannel}). 
This monotonicity relation can in turn be proven by showing $\dv{\Delta}{\mu} \lessgtr 0$ for
$\mu \gtrless \frac12$
by exploiting Eq.~(\ref{eq:dDeltadH}).
For ease of calculation, we substitute $\mu \mapsto \frac{1+\alpha}2$.
Thus, we accomplish our aim of proving $\dv{\Delta}{\mu} \lessgtr 0$ when
$\mu \gtrless \frac12$ by instead proving that
\begin{equation}
\sgn\left(\dv{\Delta}{\alpha}\right) = -\sgn(\alpha)
\end{equation}
for $-1 < \alpha < 1$.

First we derive an expression for $\dv{\Delta}{\alpha}$.
A tedious-but-straightforward calculation starting from
\cref{eq:twobitdiscordmu} reveals that
\begin{align}
\label{eq:dDeltadalpha}
 \dv{\Delta}{\alpha}
 = & \left( p_{00} - p_{01} + p_{10} - p_{11} \right)
     \log \left( \frac{
        1 + \alpha \left( p_{00} - p_{01} + p_{10} - p_{11} \right)
       }{
        1 - \alpha \left( p_{00} - p_{01} + p_{10} - p_{11} \right)
       }
     \right) \nonumber \\
 &- \left( p_{00} - p_{01} \right)
     \log \left( \frac{
        1 + \alpha \left(
          \frac{p_{00} - p_{01}}{p_{00} + p_{01}}
        \right)
       }{
        1 - \alpha \left(
          \frac{p_{00} - p_{01}}{p_{00} + p_{01}}
        \right)
       }
     \right)
 -   \left( p_{10} - p_{11} \right)
     \log \left( \frac{
        1 + \alpha \left(
          \frac{p_{10} - p_{11}}{p_{10} + p_{11}}
        \right)
       }{
        1 - \alpha \left(
          \frac{p_{10} - p_{11}}{p_{10} + p_{11}}
        \right)
       }
     \right).
\end{align}
If we assign
\begin{align}
\gamma_0 &\gets \frac{p_{00} - p_{01}}{p_{00} + p_{01}},\nonumber\\
\gamma_1 &\gets \frac{p_{10} - p_{11}}{p_{10} + p_{11}},\nonumber\\
w_0 &\gets p_{00} + p_{01},\nonumber\\
w_1 &\gets p_{10} + p_{11},\nonumber\\
f_\alpha (x) &\gets x \log \left( \frac{1 + \alpha x}{1 - \alpha x} \right),
\end{align}
we can rewrite Eq.~(\ref{eq:dDeltadalpha}) as 
\begin{equation}
 \dv{\Delta}{\alpha}
 = f_\alpha \left( w_0 \gamma_0 + w_1 \gamma_1 \right) - w_0 f_\alpha (\gamma_0) - w_1 f_\alpha (\gamma_1).
\end{equation}
Noting that $-1 < \gamma_{0,1} < 1$ and $0 < w_{0,1} < 1$
when none of $p_{00}, p_{01}, p_{10}, p_{11}$ are equal
to zero, we have $\dv{\Delta}{\alpha}$ is positive/negative/zero
when $f_\alpha$ is concave/convex/linear on the interval $(-1, 1)$.
If one or more of $p_{00}, p_{01}, p_{10}, p_{11}$ are zero,
we can have the logarithm argument become of the 0/0 type,
in which case we apply l'H\^{o}pital's rule to obtain the limit.

We now show that $f_\alpha$ is concave/linear/convex depending on
whether $\sgn\alpha=-1, 0, 1$. It is easy to see that $f_0 \equiv 0$,
which is a linear function, so we focus instead on the case $\alpha \neq 0$.
In that case, the identity $\log\left(\frac{1+y}{1-y}\right) = \arctanh y$
shows $f_\alpha (x) = \frac{1}{2\alpha} \times g(\alpha x)$,
where $g(y) = y \arctanh y$. Hence we need only show that $g(y)$
is convex for $-1 < y < 1$. 
This convexity can be seen directly from the
Maclaurin series for $\arctanh$,
\begin{equation}
    g (y) = \sum_{k=1}^\infty \frac{y^{2k+2}}{2k+1},
\end{equation}
which is a sum of convex functions and hence is convex.
Thus, $f_\alpha$ is convex/concave depending on the sign of $\alpha$.
Hence, $\dv{\Delta}{\alpha}$ has the appropriate sign;
hence, $\dv{\Delta}{H}$ is monotonic as required.

\section{Discussion}
\label{sec:discussion}
We have developed a full classical (i.e., non-quantum) theory for channel discord,
which establishes the meaning of nonzero discord
in the context of stochastic information theory.
Although classical discord and stochastic information theory were introduced in 2015~\cite{GdOS15},
key notions were sketched rather than fully developed.
Here we have given a detailed theory of non-quantum discord all the way from the context of a three-party protocol with noisy measurement to building in Hadamard notation for making expressions clear and elegant to the unprecedented connection between average {channel}  discord, average {channel} distortion and entropy of the noisy measurement (noisy channel), including monotonic relations between them, which thereby show that {channel}  discord,
in a classical setting, is a form of channel distortion arising due to one party's noisy measurement.

Scatter plots for average discord vs average distortion in Figs.~\ref{fig:ave_results-bipartite}
and~\ref{fig:ave_results_CondiPure}
show this monotonic relation between average discord and average distortion and, through the heat maps,
also the monotonicity between average distortion and channel entropy.
These results are purely numerical but show a simple quadratic relationship
for two choices of message sizes.
A general mathematical relation connecting average discord to average distortion is beyond the scope of this work
but we derive mathematical relations for the two-bit case both to illustrate how analytical results can be obtained and also to lend support to our conjectures based on numerical analysis;
our analysis has focused on developing the protocol,
making clear the problem,
defining appropriate quantities
and tackling numerically.
Mathematically proving the general case is challenging so we instead solve the special two-bit case,
i.e., the case that each of Alice and Bob hold one bit and Bob's measurement is noisy,
and there we prove that channel discord is a monotonic function of channel entropy.
Thus, the numerical results are backed up by a closed-expression analysis,
and, furthermore,
this analysis points the way to general proofs,
likely using the Hadamard calculus elaborated in Appendix~\ref{sec:hadamardcalculus}.

The plots in~\S\ref{sec:results} display a high level of scatter for medium-entropy cases and much less scatter for low- and high-entropy cases.
Although the spread is large,
monotonicity and quadratic scaling is clearly evident in these plots and the root-mean-squared error (RMSE) of each plot hovering around 0.1 is testament to the quality of the quadratic fit and hence the inference of monotonicity.

Our analysis has focused only on noise represented by doubly stochastic maps,
which correspond to permutahedrons.
In other words, we have concentrated on noise that would arise from random permutations of classical message measurements correspondingly measuring   some messages as different messages incorrectly.
In the spirit of quantum discord,
our average discord and average distortion calculations are built on minimizing over all such permutations.
Future work should involve generalization from doubly stochastic to stochastic maps;
in the quantum context,
this generalization would be akin to extending from completely positive trace-preserving mappings
to completely positive mappings.
\section{Conclusions}
\label{sec:conclusions}
Discord has emerged as one of the most significant quantum resources~\cite{Merali2011}
but not without controversy~\cite{Ferraro2010,GdOS15}.
Separating quantum and non-quantum aspects of discord is vital to determine quantum resourcefulness and otherwise for discord.
The connection between state discord and entanglement is known,
but discord for channels has been unexplored under the treatment of noisy measurement as being manifested by a noisy channel.
Here we establish and elucidate connections between classical channel discord and channel distortion and entropy.
To this end,
we have developed a protocol,
mathematical framework and numerical analysis of average channel discord,
with averaging being over random shared message states
(for random initial joint distributions and, to check consistency, over conditionally pure states)
and over random doubly stochastic channels representing noisy measurement by one party.
Note that we have 
defined classical channel discord
to quantify channel `fluctuations'
(i.e., `noise'),
in contrast to earlier work on  classical discord,
which quantifies fluctuations for a state~\cite{GdOS15}.
Our notion of classical channel discord then leads to our numerical demonstration of monotonicity  between classical (channel) discord and channel distortion.
Thus, our results show numerically that average discord,
in the non-quantum setting,
is equivalent to average distortion of a channel
with channel distortion based on total-variation distance.
Furthermore, we show numerically that this distortion measure is monotonic in channel entropy,
which builds confidence that total-variation distance is a reasonable way to quantify distortion.

Akin to quantum discord,
we have incorporated minimization of average discord and average distortion over all permutations,
which permutations referring to permuting messages.
The identity permutation corresponds to reading each message correctly and other permutations cause some messages to be read as other messages.
Given that the noisy measurement is modelled as a doubly stochastic channel,
which is a permutahedron,
the idea of minimizing over all permutations is to identify which permutation minimizes {channel}  discord and minimizes {channel} distortion averaged over all states.
This minimization is key to connecting our notion of classical distortion to the quantum version.

We have created a full framework for studying the connection between average discord and average distortion for a noisy channel and have shown numerically a monotonic relation between the two.
This monotonic relation is satisfying as we can now regard discord,
in the classical setting,
as an alternative measure of channel distortion,
manifested as noisy readout by one party.
We augment our numerical analysis of channel discord vs channel distortion
by studying the two-bit example analytically and provide results for discord corresponding to certain special but important two-bit states.
We can see that analytic methods are challenging, even in the two-bit case,
but our analytical study shows a path forward for general analytical work,
which benefits from using the Hadamard calculus we discuss in Appendix~\ref{sec:hadamardcalculus}.

\section*{Acknowledgements}
This work has been supported by the Australian Research Council (ARC) via the Centre of Excellence in Engineered Quantum Systems (EQuS) project number CE110001013.  BCS appreciates financial support from  NSFC (Grant No.\ 11675164).
The authors appreciate useful discussion with Si-Hui Tan and Nigum Arshed in the early stages of this project.
The authors acknowledge University of Sydney for providing high-performance computing used to obtain early informative results. 
\appendix
\section{Hadamard Calculus}
\label{sec:hadamardcalculus}
We review the convenient Hadamard notation~\cite{bocci2016hadamard},
and introduce what we call Hadamard calculus,
which is novel but useful for studying stochastic information.
Specifically,
we explain the Hadamard product and summing over elements of a Hadamard product,
the Hadamard logarithm
and the entropy in this notation.
Also we introduce vector calculus based on Hadamard notation principles.

The Hadamard product between two rank-$t$ tensors
\begin{equation}
\label{eq:tworankt}
    \bm{a}=\left(a_{\imath_1\imath_2\dots\imath_t}\right),\,
    \bm{b}=\left(b_{\imath_1\imath_2\dots\imath_t}\right)
\end{equation}
is
\begin{equation}
\label{eq:Hadamardproduct}
    \bm{a}\circ\bm{b}
        :=\left(a_{\imath_1\imath_2\dots\imath_t} b_{\imath_1\imath_2\dots\imath_t}\right),
\end{equation}
which is simply the rank-$t$ tensor comprising element-wise products of the elements of each of the two rank-$t$ tensors in the product.
For $t=1$,
$\bm{a}$ and~$\bm{b}$ are vectors, and
$\circ$ is just the vector obtained by element-wise products of the corresponding vector elements.
We define the sum over all elements in the Hadamard product~(\ref{eq:Hadamardproduct}) by
\begin{equation}
\label{eq:odot}
\bm{a}\odot\bm{b}
    :=\sum_{\imath_1\imath_2\cdots\imath_t}
    (\bm{a}\circ\bm{b})_{\imath_1\imath_2\cdots\imath_t}
     =\sum_{\imath_1\imath_2\cdots\imath_t} a_{\imath_1\imath_2\cdots\imath_t}
     b_{\imath_1\imath_2\cdots\imath_t}.
\end{equation}
Typically,
in the literature,
$\circ$ and~$\odot$
are both employed to refer to our~$\circ$,
but here we use~$\circ$ and~$\odot$ as distinct operations as we have explained here.
The norm of a vector, matrix or tensor is
\begin{equation}
\label{eq:norm}
\norm{\bm a}
=\sqrt{\bm{a}\odot\bm{a}},
\end{equation}
which is nonnegative.

For $\bm{a}$~(\ref{eq:tworankt}) restricted by
\begin{equation}
\label{eq:restricteda}
0 < a_{\imath_1\ldots\imath_t} \leq 1,   
\end{equation}
and introducing $\mathbb{J}:=\left(1\right)$
as the tensor with every entry being~1
(in contrast to~$\mathbb{I}$ being the matrix such that only diagonals are~1 and off-diagonal elements are all~0)
of equal size to tensor~$\bm{a}$,
the element-wise logarithm is
\begin{equation}
    \log\bm{a}:=- \sum_{\ell = 1}^\infty \frac1{\ell}
    \left( \bm1 - \bm{a}\right)^{\circ \ell}
   =\left( \log a_{\imath_1\cdots\imath_t}\right).
\end{equation}
Here we use the notation $\bullet^{\circ \ell}$ to refer to the $\ell$-fold
element-wise product of the tensor $\bullet$ with itself.

For constructing conditional states,
we employ Hadamard division $\oslash$~\cite{wetzstein2012tensor}. 
Hadamard division for two same-dimensional tensors
(including vectors and matrices)
is simply their element-wise division. 
Another definition of Hadamard division 
applies for a matrix divided by a vector,
where the length of the vector equals the number of rows (or columns) of the matrix;
in this case Hadamard division of the matrix by the vector corresponds to division of the row (or column) vectors of the matrix by the elements of a vector.

\section{Probability vector and entropy}
\label{sec:probvec}
We introduce stochastic information,
which is essentially already known~\cite{cover2012elements},
but we employ the elegant notation
of Appendix~\ref{sec:hadamardcalculus} in a novel way to convey the concepts with simple,
easy-to-grasp expressions
once Hadamard arithmetic~\cite{wetzstein2012tensor} is clear.
Below we explain the concept of a stochastic-information state.
Then we have set the stage for the subsequent discussion on bipartite stochastic-information states.

Our stochastic-information construct is a distribution of messages,
with each message labelled by integer 
\begin{equation}
\label{eq:message}
    m\in[M]:=\{1,2,\dots,M\}\subset\mathbb{Z}^+
\end{equation}
so the distribution of such messages is
\begin{equation}
\label{eq:pmM}
    \bm{p}:=\left(p_m;m\in[M]\right)\in\mathbb{R}^L,\,
    p_m\geq0\forall m,
    \sum_{m=1}^Mp_m=1
\end{equation}
represented here as a probability vector~\cite{wang2007probability}. 
The probability vector can be permuted
in the sense of rearranging the probabilities of various messages.

A given permutation is represented
by some $\sigma\in S_M$,
for~$S_M$ the permutation (or symmetry) group over all~$M$ messages.
The cardinality of~$S_M$ is $M!$,
and the permutation group can be ordered in various ways,
and we adopt  reverse  lexicographical ordering~\cite{Ziabicki1992}.
A permutation of the probability vector is  represented by permutation matrix~$\Pi_\sigma$
for permutation
$\sigma\in M_{M}\left(\{0,1\}\right)$
(i.e., $M\times M$ matrices whose entries are only~0 or~1)
given by~\cite{najnudel2013distribution}
\begin{equation}
\label{eq:Pisigmap}
    \Pi_\sigma\bm{p}
        =\left(p_{\sigma(b)}\right)\,
            \forall\bm{p}
\end{equation}
so~$\Pi_\sigma$ contains exactly one entry of~1 in each row or column and the rest of the entries are 0.
We represent the sequence of all permutation matrices by the vector
\begin{equation}
\label{eq:Pivector}
    \bm{\Pi}
        :=\left(\Pi_\sigma;
            \sigma\in S_M\right)
\end{equation}
with the sequence of~$\sigma$
drawn from the permutation group
in reverse lexicographical order.

A specific message of interest is a versor,
which is a vector whose entries all zero except one element whose entry is one~\cite{wrede2010advanced}. 
Any permutation of a versor corresponds to just replacing a given message by a new message
so a permutation of a versor is just another versor.
A versor is equivalent to a stochastic-information state
$p_m=\delta_{m\check{m}}$ for~$\check{m}$ the given message.
We write this versor,
corresponding to specific message~$\check{m}$,
as~$\bm{\delta}_{\check{m}}$.
Versors form a basis for stochastic-information states,
which we call the message basis.
A permutation matrix~(\ref{eq:Pisigmap})
is actually a tensor product of a versor and a coversor,
with a coversor defined to be a covector version of a versor.

The Cartesian product of versors forms a basis for bipartite stochastic-information states,
from which a joint distribution can be constructed.
Suppose the two parties, Alice and Bob,
each hold an information state
$\bm{\delta}^\text{A}_{\check{m}}$
and~$\bm{\delta}^\text{B}_{\check{m}'}$,
respectively,
where we use superscripts~$^\text{A}$ and~$^\text{B}$
to denote who owns which of the vector spaces in the tensor product.
Alice's and Bob's joint state is the bipartite versor
\begin{equation}
\label{eq:bipartiteversor}
\bm{\delta}^\text{AB}_{mm'}
    :=\bm{\delta}^\text{A}_m
        \bm{\delta}^\text{B}_{m'}
        \in M_{M^\text{A}\times M^\text{B}}\left(\{0,1\}\right),
\end{equation}
which is pure over the Cartesian product.
For $M^\text{A}=M^\text{B}$,
the state represented by the
$M^\text{A}\times M^\text{B}$
identity matrix~$\mathds1/\sqrt{M}$,
for~$M:=M^\text{A}M^\text{B}$,
corresponds to all messages to Alice and Bob being identical but all message instances are equally likely.
Thus, the conditional entropy of this mixture of states is zero whereas the entropy of this set of messages pairs is maximum;
i.e.,
the maximum entropy is~$\log M^\text{A}$.
\section{Quantum discord}
\label{sec:quantumdiscord}
We now explain quantum discord
at a high level instead of delving into a full mathematical description of quantum discord,
which requires Hilbert space.
We begin with a brief discussion of the history of quantum discord,
both theoretical and experimental,
and typical interpretations of what quantum discord means such as quantum correlations and resources.

The concept of quantum discord was proposed
to separate  total
correlations of a bipartite quantum state into purely quantum and classical parts~\cite{HV01,OZ01}.
Quantum discord per se is the difference between two classically identical expressions for mutual information but adapted for a quantum system and was described
as a measure of quantumness of correlations. 
Discord appeared to be a more general way of quantifying quantum correlations~\cite{Ferraro2010,vedral2017foundations,Egloff2018Of,chitambar2019quantum}
compared to entanglement
as vanishing entanglement does not
ensure vanishing discord, and the absence of entanglement does not imply classicality. 
Operationally, quantum discord has been interpreted in the context of quantum state merging, if pertinent prior information is discarded~\cite{Madhok2011,Cavalcanti2011},
the quantum-classical separation associated with discord has been cast in terms of negative conditional entropy~\cite{GdOS15}.
The value of discord as a quantum resource has been debated vigorously,
sometimes as a powerful quantum resource~\cite{Merali2011}
but also critically,
for example that pure states of nonzero discord have zero measure~\cite{Ferraro2010}
and that nonzero discord is classically explainable if and only if entanglement is not present,
as discussed in Remark~4 of Ref.~\cite{GdOS15}.
Discord has been generalised for different types of quantum measurements~\cite{xu2011generalizations}, for R\'{e}nyi entropy~\cite{hou2014quantum,Bellomo2014}
and for higher dimension~\cite{TimByrnes2020}.

Quantum discord has been experimentally studied in optical systems in which quantum discord was shown to be a resource for quantum remote-state preparation,
specifically showing that separable states with non-zero quantum discord can outperform entangled states~\cite{dakic2012quantum}.
In continuous-variable Gaussian optics,
the experimental Gaussian quantum discord has been studied for a two-mode squeezed thermal state~\cite{blandino2012homodyne}.  
Experimentally encoding information within the discordant correlations of two separable Gaussian states shows that
bipartite discord can be consumed to encode information that is only accessible by coherent quantum interactions~\cite{gu2012observing}.
A flexible two-photon setup has realized a three-qubit system with programmable degrees of initial correlations, measurement interaction, and characterization
processes, 
thereby yielding the demonstration that local observation in an activation protocol
for converting discord 
into distillable entanglement~\cite{adesso2014experimental}.  A trapped-ion experiment has shown that quantum discord inference of open-system dynamics detects system-environment quantum correlations without 
accessing the environment~\cite{gessner2014local}.

In contrast to stochastic information states,
which are probability distributions~(\ref{eq:pmM})
or joint distributions for the bipartite case~(\ref{eq:pAB}), 
the quantum state is a trace-class bounded completely positive operator~$\rho$ on Hilbert space~$\mathscr H$,
or on the tensor product $\mathscr{H}\otimes\mathscr{H}$
for the bipartite case~\cite{nielsen2002quantum}.
The quantum state's entropy is
$H(\rho)=-\operatorname{tr}\left(\rho\log\rho\right)$ 
for~tr the trace operation.
In quantum information theory,
measurement is described by positive operator-valued measures~\cite{nielsen2002quantum},
but, for quantum discord,
only projective-valued measures $\{P\}$~\cite{nielsen2002quantum},
which are self-adjoint projections on~$\mathscr H$,
are considered.
Each projective-valued measure~$P$
comprises a set of projective operators~$P_\imath$
with~$P_\imath P_\jmath=P_\imath\delta_{\imath\jmath}$.
Measurement of a state yields a real-valued outcome,
and the state is subsequently described by the $j^\text{th}$ projection~$P_j\rho$ corresponding to that outcome. 
This projection is expressed as a conjugation of $\rho$ in the literature~\cite{OZ01},
but this way of expressing is superfluous for our purposes and hence not employed.

The conditional quantum state,
first defined by Cerf and Adami~\cite{CA97},
is now typically defined as~\cite{OZ01}
\begin{equation}
\label{eq:rhoj}
    \rho_j^{\text{A}|\text{B}}
        :=\left(\mathds1\otimes P_j\right)\rho^\text{AB}\new{\left(\mathds1\otimes P_j\right)}/p_j,\;
    p_j:=\operatorname{tr}\left(\left(\mathds1
        \otimes P_j\right)\rho^\text{AB}\new{\left(\mathds1\otimes P_j\right)}\right)
\end{equation}
with
$p_j$
the probability of Bob obtaining $j^\text{th}$ outcome
after he has applied projection-valued measures (PVM)~$P$.
The conditional quantum entropy~\cite{CA97} is \begin{equation}
\label{eq:condqentropy}
    H^{\text{A}|\text{B}}=\sum_jp_jH_j^{\text{A}|\text{B}},\;
    H_j^{\text{A}|\text{B}}\left(\rho^\text{AB}\right):=H\left(\rho_j^{\text{A}|\text{B}}\right),
\end{equation}
which is a probability-weighted average of conditional entropy.

Mutual information~$I^\text{A;B}(\rho^\text{AB})$
is the same as for Eq.~(\ref{eq:I})
except that~$H$ now corresponds to the quantum entropy,
and the last term is replaced by
$H^\text{A:B}\mapsto H\left(\rho^\text{AB}\right)$.
The alternative mutual information is
\begin{equation}
\label{eq:altmutinf}
    J^{\text{A};\text{B}}
        =H^\text{A}
            -\sup_{P^\text{B}}
                H^{\text{A}|\text{B}},\;
    H^\text{A}
     :=H\left(\rho^\text{A}\right),\,
    \rho^\text{A}
        =\operatorname{tr}_\text{B}\rho^\text{AB},
\end{equation}
which is the supremum over all Bob's PVMs.
Quantum discord,
analogous to classical discord~(\ref{eq:classicaldiscorddef}),
which was actually defined later than quantum discord~\cite{GdOS15},
is
\begin{equation}
\label{eq:quantumdiscord}
    \Delta^{\text{A};\text{B}}\left(\rho^\text{AB}\right)
    :=I^{\text{A};\text{B}}\left(\rho^\text{AB}\right)-J^{\text{A};\text{B}}\left(\rho^\text{AB}\right),
\end{equation}
which quantifies the difference between the two mutual information quantities for quantum discord.

Quantum discord is interpreted as correlations that remain 
after classical correlations are subtracted from total correlation 
and recognised to quantify the non-classical correlations 
in a quantum system, including entanglement and therefore identified as a quantum resource~\cite{chitambar2019quantum}.
The primary feature encapsulated by its quantum property 
is how a state is affected by local measurements and
 seen as a form of classical 
correlation aided with quantum coherence (superpositions) at the level of individual subsystems~\cite{Ferraro2010,vedral2017foundations,WISEMAN2013361}.  
The presence of discord in quantum computing protocols~\cite{Lanyon2008, datta2008}, 
motivates the assertion that discord is a quantum resource,
operationalized by state merging~\cite{Horodecki2007},
which can deliver the quantum advantage.
\bibliographystyle{unsrt}
\bibliography{discord_distortion}

\begin{thebibliography}{10}

\bibitem{HV01}
L.~Henderson and V.~Vedral.
\newblock Classical, quantum and total correlations.
\newblock {\em J. Phys. A}, 34(35):6899--6905, aug 2001.

\bibitem{OZ01}
Harold Ollivier and Wojciech~H. Zurek.
\newblock Quantum discord: a measure of the quantumness of correlations.
\newblock {\em Phys. Rev. Lett.}, 88:017901, Dec 2001.

\bibitem{brunner2014bell}
Nicolas Brunner, Daniel Cavalcanti, Stefano Pironio, Valerio Scarani, and
  Stephanie Wehner.
\newblock Bell nonlocality.
\newblock {\em Rev. Mod. Phys.}, 86:419--478, Apr 2014.

\bibitem{Horodecki2007}
Micha{\l} Horodecki, Jonathan Oppenheim, and Andreas Winter.
\newblock Quantum state merging and negative information.
\newblock {\em Commun. Math. Phys.}, 269(1):107--136, Jan 2007.

\bibitem{Lanyon2008}
B.~P. Lanyon, M.~Barbieri, M.~P. Almeida, and A.~G. White.
\newblock Experimental quantum computing without entanglement.
\newblock {\em Phys. Rev. Lett.}, 101:200501, Nov 2008.

\bibitem{datta2008}
Animesh Datta, Anil Shaji, and Carlton~M. Caves.
\newblock Quantum discord and the power of one qubit.
\newblock {\em Phys. Rev. Lett.}, 100:050502, Feb 2008.

\bibitem{Merali2011}
Zeeya Merali.
\newblock Quantum computing: the power of discord.
\newblock {\em Nature}, 474(7349):24--26, 2011.

\bibitem{GdOS15}
Vlad Gheorghiu, Marcos~C. de~Oliveira, and Barry~C. Sanders.
\newblock Nonzero classical discord.
\newblock {\em Phys. Rev. Lett.}, 115:030403, Jul 2015.

\bibitem{nielsen2002quantum}
Michael~A. Nielsen and Isaac Chuang.
\newblock {\em {Q}uantum {C}omputation and {Q}uantum {I}nformation}.
\newblock Cambridge University Press, Cambridge, 2011.

\bibitem{shannon1948mathematical}
Claude~E Shannon.
\newblock A mathematical theory of communication.
\newblock {\em Bell Syst. Tech. J.}, 27(3):379--423, 1948.

\bibitem{cover2012elements}
Thomas~M. Cover and Joy~A. Thomas.
\newblock {\em Elements of {I}nformation {T}heory}.
\newblock John Wiley \& Sons, 2012.

\bibitem{milonni1976}
P.~W Milonni.
\newblock Semiclassical and quantum-electrodynamical approaches in
  nonrelativistic radiation theory.
\newblock {\em Phys. Rep.}, 25(1):1--81, 1976.

\bibitem{Gasbarri2018}
G.~Gasbarri and L.~Ferialdi.
\newblock Stochastic unravelings of non-markovian completely positive and
  trace-preserving maps.
\newblock {\em Phys. Rev. A}, 98:042111, Oct 2018.

\bibitem{budish2009implementing}
Eric Budish, Yeon-Koo Che, Fuhito Kojima, and Paul Milgrom.
\newblock Implementing random assignments: a generalization of the birkhoff-von
  neumann theorem.
\newblock In {\em In 2009 Cowles Summer Conference}, 2009.

\bibitem{sinkhorn1964relationship}
Richard Sinkhorn.
\newblock A relationship between arbitrary positive matrices and stochastic
  matrices.
\newblock {\em Can. J. Math.}, 18:303--306, 1966.

\bibitem{johnson1981row}
Charles~R Johnson.
\newblock Row stochastic matrices similar to doubly stochastic matrices.
\newblock {\em Linear Multilinear Algebra}, 10(2):113--130, 1981.

\bibitem{LevinDA2009}
D.~A. Levin and Y.~Peres.
\newblock {\em Markov {C}hains and {M}ixing {T}imes}.
\newblock MBK. American Mathematical Society, 2017.

\bibitem{rezaei2012optimal}
Farzad Rezaei, Charalambos~D. Charalambous, and N.~U. Ahmed.
\newblock Optimal control of uncertain stochastic systems subject to total
  variation distance uncertainty.
\newblock {\em SIAM J. Control Optim.}, 50(5):2683--2725, 2012.

\bibitem{charalambous2014extremum}
Charalambos~D Charalambous, Ioannis Tzortzis, Sergey Loyka, and Themistoklis
  Charalambous.
\newblock Extremum problems with total variation distance and their
  applications.
\newblock {\em IEEE Trans. Autom. Control}, 59(9):2353--2368, 2014.

\bibitem{gulati2006Testing}
Sneh Gulati and Jie Mi.
\newblock Testing for scale families using total variation distance.
\newblock {\em J. Stat. Comput. Simul.}, 76(9):773--792, 2006.

\bibitem{Ferraro2010}
A.~Ferraro, L.~Aolita, D.~Cavalcanti, F.~M. Cucchietti, and A.~Ac\'{\i}n.
\newblock Almost all quantum states have nonclassical correlations.
\newblock {\em Phys. Rev. A}, 81:052318, May 2010.

\bibitem{bocci2016hadamard}
Cristiano Bocci, Enrico Carlini, and Joe Kileel.
\newblock Hadamard products of linear spaces.
\newblock {\em J. Algebra}, 448:595--617, 2015.

\bibitem{wetzstein2012tensor}
Gordon Wetzstein, Douglas Lanman, Matthew Hirsch, and Ramesh Raskar.
\newblock Tensor displays: compressive light field synthesis using multilayer
  displays with directional backlighting.
\newblock {\em ACM Transactions on Graphics}, 31:1--11, 2012.

\bibitem{wang2007probability}
Xing~M Wang.
\newblock Probability bracket notation, probability vectors, markov chains and
  stochastic processes.
\newblock {\em Preprint, arXiv:cs/0702021}, Feb 2007.

\bibitem{Ziabicki1992}
Andrzej Ziabicki.
\newblock The theory of ordering lexicographic entries: principles, algorithms
  and computer implementation.
\newblock {\em Comput. Hum.}, 26(2):119--137, 1992.

\bibitem{najnudel2013distribution}
Joseph Najnudel and Ashkan Nikeghbali.
\newblock The distribution of eigenvalues of randomized permutation matrices.
\newblock volume~63, pages 773--838. Association des Annales de l'institut
  Fourier, 2013.

\bibitem{wrede2010advanced}
L.~H. Loomis and S.~Sternberg.
\newblock {\em Advanced {C}alculus: {R}evised Edition}.
\newblock World Scientific, 2014.

\bibitem{vedral2017foundations}
Vlatko Vedral.
\newblock Foundations of quantum discord.
\newblock In {\em Lectures on {G}eneral {Q}uantum {C}orrelations and their
  {A}pplications}, pages 3--7. Springer, 2017.

\bibitem{Egloff2018Of}
Dario Egloff, Juan~M. Matera, Thomas Theurer, and Martin~B. Plenio.
\newblock Of local operations and physical wires.
\newblock {\em Phys. Rev. X}, 8:031005, Jul 2018.

\bibitem{chitambar2019quantum}
Eric Chitambar and Gilad Gour.
\newblock Quantum resource theories.
\newblock {\em Rev. Mod. Phys.}, 91:025001, Apr 2019.

\bibitem{Madhok2011}
Vaibhav Madhok and Animesh Datta.
\newblock Interpreting quantum discord through quantum state merging.
\newblock {\em Phys. Rev. A}, 83:032323, Mar 2011.

\bibitem{Cavalcanti2011}
D.~Cavalcanti, L.~Aolita, S.~Boixo, K.~Modi, M.~Piani, and A.~Winter.
\newblock Operational interpretations of quantum discord.
\newblock {\em Phys. Rev. A}, 83:032324, Mar 2011.

\bibitem{xu2011generalizations}
Jianwei Xu.
\newblock Generalizations of quantum discord.
\newblock {\em J. Phys. A}, 44(44):445310, 2011.

\bibitem{hou2014quantum}
Xi-Wen Hou, Zhi-Peng Huang, and Su~Chen.
\newblock Quantum discord through the generalized entropy in bipartite quantum
  states.
\newblock {\em Eur. Phys. J. D}, 68(4):87, 2014.

\bibitem{Bellomo2014}
Guido Bellomo, Angelo Plastino, Anna~P. Majtey, and Angel~R. Plastino.
\newblock Comment on ``{Q}uantum discord through the generalized entropy in
  bipartite quantum states".
\newblock {\em Eur. Phys. J. D}, 68(11):337, 2014.

\bibitem{TimByrnes2020}
Chandrashekar Radhakrishnan, Mathieu Lauri\`ere, and Tim Byrnes.
\newblock Multipartite generalization of quantum discord.
\newblock {\em Phys. Rev. Lett.}, 124:110401, Mar 2020.

\bibitem{dakic2012quantum}
Borivoje Daki{\'c}, Yannick~Ole Lipp, Xiaosong Ma, Martin Ringbauer, Sebastian
  Kropatschek, Stefanie Barz, Tomasz Paterek, Vlatko Vedral, Anton Zeilinger,
  {\v{C}}aslav Brukner, et~al.
\newblock Quantum discord as resource for remote state preparation.
\newblock {\em Nature Phys.}, 8(9):666--670, 2012.

\bibitem{blandino2012homodyne}
R\'emi Blandino, Marco~G. Genoni, Jean Etesse, Marco Barbieri, Matteo G.~A.
  Paris, Philippe Grangier, and Rosa Tualle-Brouri.
\newblock Homodyne estimation of gaussian quantum discord.
\newblock {\em Phys. Rev. Lett.}, 109:180402, Nov 2012.

\bibitem{gu2012observing}
Mile Gu, Helen~M Chrzanowski, Syed~M Assad, Thomas Symul, Kavan Modi, Timothy~C
  Ralph, Vlatko Vedral, and Ping~Koy Lam.
\newblock Observing the operational significance of discord consumption.
\newblock {\em Nature Phys.}, 8(9):671--675, 2012.

\bibitem{adesso2014experimental}
Gerardo Adesso, Vincenzo D'Ambrosio, Eleonora Nagali, Marco Piani, and Fabio
  Sciarrino.
\newblock Experimental entanglement activation from discord in a programmable
  quantum measurement.
\newblock {\em Phys. Rev. Lett.}, 112:140501, Apr 2014.

\bibitem{gessner2014local}
M.~Gessner, M.~Ramm, T.~Pruttivarasin, A.~Buchleitner, H.-P. Breuer, and
  H.~H{\"a}ffner.
\newblock Local detection of quantum correlations with a single trapped ion.
\newblock {\em Nature Phys.}, 10(2):105--109, Feb 2014.

\bibitem{CA97}
N.~J. Cerf and C.~Adami.
\newblock Negative entropy and information in quantum mechanics.
\newblock {\em Phys. Rev. Lett.}, 79:5194--5197, Dec 1997.

\bibitem{WISEMAN2013361}
Howard~M. Wiseman.
\newblock Quantum discord is {B}ohr's notion of non-mechanical disturbance
  introduced to counter the einstein–podolsky–rosen argument.
\newblock {\em Ann. Phys.}, 338:361--374, 2013.

\end{thebibliography}
\end{document}